\documentclass[11pt]{article}
\pdfoutput=1
\usepackage{jheppub}
\usepackage{amsmath}
\usepackage{bm}
\usepackage{slashed}
\usepackage{graphicx}
\usepackage{tabularx}
\usepackage{diagbox}
\usepackage{slashed}

\newlength{\dummysp}
\newcommand{\beq}{\begin{eqnarray}}
\newcommand{\eeq}{\end{eqnarray}}

\newcommand{\gappeq}{\mathrel{\rlap {\raise.5ex\hbox{$>$}}
{\lower.5ex\hbox{$\sim$}}}}
\newcommand{\lappeq}{\mathrel{\rlap{\raise.5ex\hbox{$<$}}
{\lower.5ex\hbox{$\sim$}}}}

\newcommand{\ben}{\begin{enumerate}}
\newcommand{\een}{\end{enumerate}}

\newcommand{\bit}{\begin{itemize}}
\newcommand{\eit}{\end{itemize}}

\def\[{\left [}
\def\]{\right ]}
\def\({\left (}
\def\){\right )}

\title{Self-conjugate QCD}
   
\author{Mohamed M. Anber}
\affiliation{Department of Physics, Lewis $\&$ Clark College, Portland, OR 97219, USA}
\emailAdd{manber@lclark.edu}

\abstract{We carry out a systematic study of $SU(6)$ Yang-Mills theory endowed with fermions in the adjoint and $3$-index antisymmetric mixed-representation. The fermion bilinear in the $3$-index antisymmetric representation vanishes identically, which leads to interesting new phenomena.  We first study the theory on a small circle, i.e., on  $\mathbb R^3\times \mathbb S^1_L$, employing symmetry-twisted boundary conditions and semi-classical techniques.  We find that the ground state is $3$-fold degenerate, which can be explained as a consequence of a $1$-form/$0$-form mixed 't Hooft anomaly. In addition, the theory  may admit massless bosonic and fermionic degrees of freedom, depending on the number of flavors, and confines the electric probes in the infrared. Empowered by 't Hooft anomaly matching conditions along with the $2$-loop $\beta$-function, we  further examine the  possible infrared symmetry realizations on $\mathbb R^4$ for various number of adjoint and $3$-index antisymmetric fermions. The infrared theory is either a conformal field theory, which is expected  for  a large number of flavors, or it is confining with or without chiral symmetry breaking.  In a few cases, we are able to give enough evidence for  adiabatic continuity between the small- and large-circle limits.}

\begin{document}

\maketitle

\flushbottom

\section{Introduction}

Supersymmetry is a leading candidate for physics beyond the Standard Model.  These theories enjoy a holomorphic structure that greatly constraints their infrared (IR) dynamics, which  is particularly helpful when they become strongly coupled. However, we still do not know whether nature will make use of supersymmetry or some other class of theories at energies above the TeV scale. Given the set of the renormalizable and ultraviolet (UV) complete gauge theories, it would be very desirable to have a systematic study of important subsets of these theories and their strong dynamics.  For this reason,  any technique that can shed light on the strong dynamics beyond supersymmetry is greatly appreciated. 

A  powerful  tool that provides a  handle on understanding  gauge theories beyond perturbation theory is 't Hooft anomalies \cite{tHooft:1980xss}. Given a global symmetry ${\cal G}$, we may try to gauge it by turning on a background  field of ${\cal G}$. The obstruction to gauging ${\cal G}$ is called a 't Hooft anomaly. The anomaly is a renormalization group invariant, and therefore, it encodes information about the theory at all scales. The matching of 't Hooft anomalies in the IR put sever constraints on the possible realizations of its global symmetries. Recently, it has been realized that in addition to the anomalies of the $0$-form symmetries, which act on local fields, one may   turn on background gauge fields of higher-form symmetries \cite{Gaiotto:2014kfa}. The obstruction to gauging these symmetries, hence higher-form 't Hooft anomalies \cite{Gaiotto:2017yup},  work as additional constraints on the possible realizations of the IR dynamics, see \cite{Tanizaki:2018wtg,Benini:2018reh,Choi:2018tuh,Komargodski:2017smk,Shimizu:2017asf,Komargodski:2017dmc,Kikuchi:2017pcp,Aitken:2018kky,Tanizaki:2018xto,Sulejmanpasic:2018upi,Tanizaki:2017mtm,Anber:2018xek,Anber:2018jdf,Karasik:2019bxn,Cordova:2019jnf,Cordova:2019uob,Misumi:2019dwq,Nishimura:2019umw} for recent advances. 

Another approach that has gained momentum over the past decade is to compactify a strongly coupled theory on a circle $\mathbb S_L^1$.  We say that the theory lives on $\mathbb R^3\times \mathbb S^1_L$. If we take the circle size to be small enough, much smaller than the inverse strong-coupling scale, add deformations, and/or give fermions twisted boundary conditions on $\mathbb S_L^1$, then  the theory enters its weakly coupled regime and becomes amenable to  semi-classical analysis. Now, one can use analytical tools to study the vacuum and sort out the IR dynamics. The twisted boundary conditions turn the thermal partition function into a graded-state sum and lead to a cancelation among  the  excited states,  retaining mainly the ground state of the system. In other words, the graded partition function {\em quantum distills} the ground state \cite{Dunne:2018hog}.  The caveat, however, is that there is a possibility that the theory  exhibits a phase transition on the way of compactification from $\mathbb R^4$ to $\mathbb R^3\times \mathbb S^1_L$. How to prevent the phase transition and allow for an adiabatic continuity as we change the circle size is still an open question.    Nevertheless, this procedure is invaluable to making a few aspects of strongly coupled theories manifest in a weakly coupled regime. This includes, but not limited to, studies of the confinement mechanism, chiral symmetry breaking, string spectrum, entanglement entropy, and many other phenomena, see \cite{Anber:2015kea,Cherman:2016hcd,Anber:2011gn,Poppitz:2012sw,Anber:2012ig,Unsal:2008ch,Anber:2013doa,Anber:2014lba,Dunne:2016nmc,Anber:2014sda,Anber:2018ohz,Aitken:2017ayq} for a non-comprehnsive list of applications. 

There has been only a few studies, see \cite{Tanizaki:2017qhf,Dunne:2018hog,Hongo:2018rpy,Anber:2018tcj,Misumi:2019dwq}, that attempted to combine both methods, 't Hooft anomalies and compactification, to maximize our learning about strongly coupled theories. The purpose of this work is to further examine the interplay between both methods and use them to shed light on gauge theories endowed with fermions in the self-conjugate representations. These are real representations of the gauge group, and hence, they are free from gauge anomalies for any number of fermions. Given a gauge group $SU(N)$, a representation is said to be self-conjugate if it satisfies\footnote{Here, we are using the Dynkin indices to label the irreducible representations. See Appendices A-E of \cite{Anber:2017pak} for the group theory convention and normalization used in this work.} $(a_1,a_2,...,a_{N-2},a_{N-1})=(a_{N-1},a_{N-2},...,a_2,a_1)$, i.e., is equal to its conjugate representation.  Famous examples include the adjoint representation and the $N$-index antisymmetric representation of $SU(2N)$.  

In this paper we study $SU(6)$ Yang-Mills theory with fermions in the adjoint and $3$-index antisymmetric mixed-representation. The reason for studying this theory is multifold. First, both representations can be used in model building beyond the Standard Model. For example, a relatively large number of fermions in both representations can push the theory towards its conformal/near-conformal window, which can have applications that address the dynamical electroweak symmetry breaking, see e.g. \cite{Dietrich:2006cm}.  Second, the fermion bilinear in the $3$-index antisymmetric representation vanishes identically. The absence of a mass term puts this theory on equal footing with chiral gauge theories. The latter have attracted a lot of attention because their IR dynamics can be richer than vector-like theories, see \cite{Poppitz:2010at} for a review and  \cite{Bolognesi:2019wfq,Ryttov:2019evz} for recent progress. Studying $SU(6)$ Yang-Mills theory with fermions in the $3$-index antisymmetric representation, in turn, may enlighten us about the vacuum structure in their cousins, chiral gauge theories.  Last but not least, there is a general consensus that one needs to learn from many examples before a complete picture of the adiabatic continuity on the circle can emerge. The fact that the theory under study has a plethora of 't Hooft anomalies, both $0$- and $1$-form anomalies, makes it a rich playground to examine the connection between the spectrum on $\mathbb R^3\times \mathbb S^1_L$ and on $\mathbb R^4$ and check for signs of adiabatic continuity in the decompactification limit. 
 
The theory on $\mathbb R^3\times \mathbb S^1_L$ will be solved via semi-classical techniques. In the absence of adjoint fermions we find that the discrete chiral symmetry is broken, the vacuum is $3$-fold degenerate, the IR spectrum is fully gapped, and the theory confines the fundamental electric probes. Adding adjoints does not change the vacuum degeneracy or the confinement of the electric charges. However, it introduces massless excitations to the theory: one bosonic and one fermionic degree of freedom\footnote{Though, a single adjoint fermion acquires a mass from the monopole operators. See the bulk of the paper for details.}. The latter transforms in the fundamental representation of the flavor group of the adjoint fermions. The observation that adding adjoint fermions does not alter the number of the ground states is ultimately tied to a mixed 't Hooft anomaly between the $1$-form $\mathbb Z_3^{(1)}$  center and the $0$-form discrete chiral symmetries, as was argued before in \cite{Yamaguchi:2018xse}. The saturation of this anomaly in the IR demands a breaking pattern of the discrete symmetries that yields $3$ degenerate ground states.  Continuous chiral symmetries, however, are intact on the small circle, thanks to the weak coupling nature of the theory. 

The story on $\mathbb R^4$ is more involved and it takes different twists depending on the number of flavors in both representations.  A large number of fermions will push the theory inside its conformal window: the theory flows to a fixed point in the IR. The situation here resembles QCD(adj) with a large number of adjoint flavors. Definitely, because the dimension of  the $3$-index antisymmetric representation is smaller than the dimension of the adjoint representation,  one can make a room inside the conformal window for a larger number of fermions in the former representation.   For a smaller number of fermions, the theory has to break its continuous chiral symmetries\footnote{It is instructive to mention that the absence of a mass term for the $3$-index antisymmetric fermions does not imply the absence of condensates. Indeed, the theory can still break its chiral symmetry via higher-order operators.}. Otherwise,  massless composite fermions will populate the vacuum. The idea is that  there are various non-trivial $0$-form 't Hooft anomalies in the UV that have to be matched in the IR by massless excitations: Goldstone bosons or composite fermions\footnote{Matching by a conformal field theory is the third, trivial, possibility.}. Which of the two options will be realized is a dynamics question that ultimately can only be answered in an experimental (lattice) setup. In both situations, the $\mathbb Z_3^{(1)}$ center symmetry is intact and the Wilson loops obey the area law. In addition, matching the $1$-form/$0$-form mixed 't Hooft anomaly requires the $3$-fold degeneracy of the ground state, exactly as in the theory on circle, or that there is an IR topological quantum field theory (TQFT) that matches the anomaly. In this work we address all these issues and give a detailed description of the different scenarios. 

If the theory on $\mathbb R^4$ flows to a conformal field theory (CFT), then strictly speaking  its spectrum is different from the one on $\mathbb R^3\times \mathbb S^1_L$. However, if the mass gap on $\mathbb R^3\times \mathbb S^1_L$ is a decreasing function of the circle size $L$ and always stays  lighter than the W-bosons for all values of $L$, then the theory is under control for all values of $0<L<\infty$. We find no evidence that this is the case for any number of adjoint or $3$-index antisymmetric flavors\footnote{It is important to emphasize that our approach in determining the conformal window is different from the works \cite{Poppitz:2009uq,Poppitz:2009tw}, where the main idea was to  employ the mass gap for gauge fluctuations as an invariant characterization of conformality versus confinement.}.

With the aid of the $2$-loop $\beta$-function, we employ 't Hooft anomaly matching conditions to give a strong evidence that Yang-Mills theory with a single fermion in the $3$-index antisymmetric representation is adiabatically continuous on $\mathbb R^3\times \mathbb S^1_L$ for all values of $L$, i.e.,  it does not experience a phase transition as we change the circle size. We also show that a theory with $2$ or more fermions  in the $3$-index antisymmetric representation has to experience a phase transition in the decompactification limit. There is also evidence that theories with a single adjoint and a single $3$-index antisymmetric fermions do not experience a phase transition in the decompactification limit.  Adding more adjoint fermions to the story makes it more complicated and no strong conclusions can be reached. However, the need for a large number of composite fermions in the IR to match all the anomalies suggest that the theory might need to break its continuous symmetries, otherwise the theory should flow to a CFT.

This paper is organized as follows. In Section \ref{Theory and formulation} we formulate our theory, enumerate its $1$-form and $0$-form symmetries, and use the $2$-loop $\beta$-function to speculate about its IR phases on $\mathbb R^4$.  In Section \ref{Semi-classical treatment} we compactify the theory on a small circle and study it using the semi-classical techniques: we analyze the monopoles, their fermion zero modes, and topological molecules. We show that the proliferation of the topological molecules give rise to a mass gap: the theory confines the electric charges. We also study the IR spectrum and  vacuum  of the theory for various number of adjoint and $3$-index antisymmetric fermions and show that the latter is always $3$-fold degenerate. In Section \ref{t Hooft anomalies and IR spectrum} we use both the $1$-form and $0$-form 't Hooft anomalies to put constraints on the IR spectrum on $\mathbb R^4$ for various number of fermions, and show that strong conclusions can be reached in some cases. We conclude our work with a brief outlook in Section \ref{outlock}.

\section{Theory and formulation}
\label{Theory and formulation}

We consider $SU(N=6)$ Yang-Mills theory with fermions in the adjoint ($G$) and $3$-index antisymmetric ($\cal R$) representations.  We denote the adjoint fermions by $\chi$, while fermions in ${\cal R}$ are denoted by $\lambda$. The Lagrangian of the theory reads
\begin{eqnarray}
{\cal L}=-\frac{1}{2g^2}\mbox{tr}_F\left[F_{MN}F^{MN}\right]+i\mbox{tr}_F\left[\bar\chi_p\bar \sigma^M\partial_M \chi^p\right]+i\mbox{tr}_F\left[\bar\lambda_q\bar \sigma^M\partial_M \lambda^q\right]\,,
\label{Lagrangian of the system}
\end{eqnarray}
where $M,N=0,1,2,3$ are the spacetime indices and $p,q$ are the flavor indices: $p=1,2,...,n_G$, $q=1,2,...,n_{\cal R}$. Throughout this work we use the normalization $\mbox{tr}_F\left[ T^aT^b \right]=\delta^{ab}$, where $\{T^a\}$ are the Lie-algebra generators. This amounts to normalizing the simple and affine roots\footnote{There are $N-1$ simple roots in $SU(N)$ group : $\bm \alpha_a$, $a=1,2,...,N-1$, while the affine roots $\bm\alpha_0$ is given by
\begin{eqnarray}
\bm \alpha_0=-\sum_{a=1}^{N-1}\bm \alpha_a\,.
\end{eqnarray} 
} as $\bm \alpha_a\cdot \bm \alpha_a=2$ for all $a=0,2,...,5$. We will also use the latin letters $i,j,k,..,$ etc. $\in \{1,2,...,N\}$ to denote the color indices of $SU(6)$. In particular, the $3$-index antisymmetric fermion is denoted by $\lambda^{i_1i_2i_3}$, where the antisymmetrization of the indices is implicitly understood. Also, the adjoint fermion has the index structure $\chi_i^j$.  Since both $G$ and ${\cal R}$ are self-conjugate (real) representations, the theory does not suffer from gauge anomalies for all values of $n_G$ and $n_{\cal R}$.

\subsection*{Symmetries}

To further study this theory, we need to find its $0$- and $1$-form global symmetries that survive the quantum corrections. We can directly read the $1$-form symmetry by compactifying the theory on a circle and examining the quantum generated effective potential. Doing so, we will also be ready to use the same potential to study the theory on a small circle using semi-classical techniques\footnote{In Section \ref{t Hooft anomalies and IR spectrum} we give a shortcut method to directly read the center symmetry of a general theory with fermions in mixed representations.}. 

To this end, we compactify one of the spacial directions on a circle $\mathbb S^1_L$ of circumference $L$ and give both sets of fermions periodic boundary conditions. The compactification results in a tower of massive Kaluza-Klein excitations of both the gauge field and fermions. Upon integrating out this tower, a potential of the holonomy, the gauge field component along the compact direction, is generated. Let us take the circle along the $x^3$-direction. Also, we can always take the holonomy along the Cartan generators in the color space\footnote{This can be achieved via a global $SU(6)$ transformation.}.  We define the dimenstionless holonomy $\bm \Phi$ as $\bm\Phi \equiv L\bm A^3$, where the boldface letters denote quantities in the Cartan directions\footnote{The Cartan space  of $SU(N)$ has dimension $N-1$.}.  Then, the potential $V(\bm\Phi)$ is given by (see \cite{Anber:2017pak} for details):
\begin{eqnarray}
V\left(\bm \Phi\right)=\frac{2}{\pi^2L^3}\sum_{p=1}^\infty\frac{\mbox{Re}\left\{n_{{\cal R}}\mbox{tr}_{{\cal R}}\left[\Omega^p\right]+(n_{G}-1)\mbox{tr}_{G}\left[\Omega^p\right]  \right\}}{p^4}\,,
\label{potential in terms of the polyakov loop}
\end{eqnarray}
where $\Omega\equiv e^{i\bm \Phi\cdot \bm H}$ is the Polyakov's loops that wraps $\mathbb S_L^1$ and $\bm H$ are the Cartan generators. 
One can further simplify this expression using the Frobenius formula \cite{Anber:2017pak,Myers:2009df,Gross:1993yt}, which gives the $G$ and ${\cal R}$ traces in terms of the fundamental trace: 
\begin{eqnarray}
\nonumber
\mbox{tr}_G\left[\Omega^p\right]&=&|\mbox{tr}_F\left[\Omega^p\right]|^2-1\,,\\
\mbox{tr}_{\cal R}\left[\Omega^p\right]&=&-\frac{1}{2}\mbox{tr}_F\left[\Omega^p\right]\mbox{tr}_F\left[\Omega^{2p}\right]+\frac{1}{3}\mbox{tr}_F\left[\Omega^{3p}\right]+\frac{1}{6}\left(\mbox{tr}_F\left[\Omega^{p}\right]\right)^3\,.
\label{using Frobenius formula}
\end{eqnarray}
From (\ref{using Frobenius formula}) one can immediately see that $V\left(\bm \Phi\right)$ is invariant under the $\mathbb Z_3$ discrete symmetry transformation $\Omega\rightarrow e^{i\frac{2\pi k}{3}}\Omega$, where $k=0,1,2$. Therefore, we conclude that the theory in hand has a $1$-form $\mathbb Z^{(1)}_3$ center symmetry. 

As a side note,  the reader should refrain from trusting (\ref{potential in terms of the polyakov loop}) for all values of $L$. As we will see in Section \ref{Semi-classical treatment}, this expression is reliable only in the regime $L\Lambda\ll1$, where $\Lambda$ is the strong coupling scale of the theory. So far, the form of $V\left(\bm \Phi\right)$ was solely used to read the center symmetry of the theory in a lucid way.

In addition to the $1$-form $\mathbb Z^{(1)}_3$ symmetry, the theory enjoys a plethora of $0$-form symmetries. The full classical symmetry of the theory is $SU(n_G)\times SU(n_{\cal R})\times U_G(1)\times U_{\cal R}(1)$. However, only a subgroup of $U_G(1)\times U_{\cal R}(1)$ is good on the quantum level. In order to determine the symmetries that survive the quantum corrections, we examine the 't Hooft vertex of a BelavinPolyakov-Schwarz-Tyupkin (BPST) instanton under the symmetry transformations. Schematically,  the 't Hooft vertex  is given by\footnote{The factors $\left(\chi\chi\right)^{\frac{n_GT(G)}{2}}$ and $(\lambda\lambda)^{\frac{n_{\cal R}T(\cal R)}{2}}$ are the fermion zero modes that dress the BPST instanton.}
\begin{eqnarray}
{\cal I}=e^{-S_I}\left(\chi\chi\right)^{\frac{n_GT(G)}{2}}(\lambda\lambda)^{\frac{n_{\cal R}T(\cal R)}{2}}\,,
\end{eqnarray}
where  $S_I=\frac{8\pi^2}{g^2}$ is the BPST instanton action and $T({\cal R})=6, T(G)=12$ are the trace operators. We take the fermions to transform as
$
\chi\rightarrow \chi e^{i\alpha}\,,
\lambda\rightarrow \lambda e^{i\beta}\,,
$
under both $U_G(1)$ and $U_{{\cal R}}(1)$, respectively. Then, we find that the BPST vertex is invariant under the subgroup $U_{G-{\cal R}}(1) \subset U_G(1)\times U_{{\cal R}}(1)$, which acts as
\begin{eqnarray}
\chi\rightarrow  e^{in_{\cal R}\alpha}\chi\,,\quad
\lambda\rightarrow  e^{-i2n_G\alpha} \lambda\,.
\end{eqnarray}
Upon further inspection, one can show that the vertex is also invariant under the two independent discrete symmetry transformations:
 \begin{eqnarray}
\chi\rightarrow e^{i\gamma_1}\chi\,,\quad
\lambda\rightarrow  e^{i\gamma_2}\lambda\,,
\end{eqnarray}
where
\begin{eqnarray}
\gamma_1=\frac{2\pi k_1}{12	n_G}\,\,\,, k_1=1,2,..., 12n_G\,, \quad \gamma_2=\frac{2\pi k_2}{6 n_{\cal R}}\,\,\, k_2=1,2,...,6n_{\cal R}\,.
\end{eqnarray}
Therefore, the theory is invariant under the additional discrete groups $\left(\mathbb Z^G_{12n_G}\subset U_G(1)\right)$\newline $\times \left(\mathbb Z^{\cal R}_{6n_{\cal R}}\subset U_{\cal R}(1)\right)$, which work on $G$ and ${\cal R}$.  Collecting everything and modding out the redundant groups, we find that the full symmetry of the theory is
\begin{eqnarray}
\frac{SU(n_G)\times \mathbb Z^{G}_{12n_G}}{\mathbb Z_{n_G}}\times \frac{SU(n_{\cal R})\times \mathbb Z^{\cal R}_{6n_{\cal R}}}{\mathbb Z_{n_{\cal R}}}\times U_{G-{\cal R}}(1)\times \mathbb Z^{(1)}_3\,.
\label{all good symmetries}
\end{eqnarray}

Our next task is to examine the realization of the symmetries in the IR. To this end, we first need to study the $\beta$-function, which encodes important information about the theory.

\subsection*{The $2$-loop $\beta$-function and IR phase structure}

The phase structure of the theory on $\mathbb R^4$ can be partially envisaged by studying the fate of the various symmetries under the renormalization group flow of the coupling constant.   The two-loop $\beta$ function of the theory is given by \cite{Caswell:1974gg,Dietrich:2006cm}
\begin{eqnarray}
\nonumber
\beta(g)&=&-\beta_0\frac{g^3}{(4\pi)^2}-\beta_1\frac{g^5}{(4\pi)^4}\,,\\
\nonumber
\beta_0&=&\frac{11}{6}C_2(G)-\frac{1}{3}T(G)n_{G}-\frac{1}{3}T({\cal R})n_{\cal R}\,,\\
\nonumber
\beta_1&=&\frac{34}{12}C_2^2(G)-\frac{5}{6}n_G C_2(G)T(G)-\frac{n_G}{2}C_2(G)T(G)-\frac{5}{6}n_{\cal R} C_2(G)T({\cal R})-\frac{n_{\cal R}}{2}C_2({\cal R})T({\cal R})\,,\\
\label{beta function}
\end{eqnarray} 
where $C_2({\cal R})=\frac{21}{2}, C_2(G)=12$ are the Casimir operators. The condition that the theory remains asymptotically free, i.e., $\beta_0>0$, is given by\footnote{We made use of the identity $T({\cal R})d(G)=C_2({\cal R})d({\cal R})$.} 
\begin{eqnarray}
n_{G}+\frac{C_2({\cal R})d({\cal R})}{420}n_{{\cal R}}<\frac{11}{2}\,,
\label{the main condition}
\end{eqnarray}
where $d({\cal R})=20, d(G)=35$ are the dimensions of the representations.  In Table \ref{classification} we list the  numbers of adjoint $n_G$ and $3$-index antisymmetric $n_{\cal R}$ fermions that render the theory asymptotically free. The theory develops fixed points (a two-loop effect)  for a certain range of $n_G$ and $n_{\cal R}$. The coupling constant at the fixed point is given by
\begin{eqnarray}
\alpha^*\equiv \frac{g_*^2}{4\pi}=-4\pi\frac{\beta_0}{\beta_1}=-\frac{4\pi \left(22-4n_G-2n_{\cal R}\right)}{408-192n_G-\frac{183}{2}n_{\cal R}}
\label{alpha star}
\end{eqnarray}
and its value for different number of adjoint and ${\cal R}$ flavors is listed in Table \ref{classification}. 

Our task now is to determine the phase of the theory deep in the IR: whether it is confining (with or without chiral symmetry breaking) or  inside the conformal window. For simplicity, we first assume that $n_{\cal R}=0$.  Then, one needs to compare $\alpha^*$ to the value of the coupling constant that triggers chiral symmetry breaking $\alpha_c$, which for adjoint fermions is given by
\begin{eqnarray}
\alpha_c=\frac{2\pi}{3C_2(G)}\,.
\label{ladder approximations value}
\end{eqnarray}
This is the value of the coupling constant (in the ladder approximation, see \cite{Appelquist:1988yc}) for which the anomalous dimension of the adjoint quark mass operator becomes unity. As we flow from high to low energies, the coupling constant keeps growing as long as $\beta_0>0$. If we reach $\alpha_c$ before hitting a fixed point, then chiral symmetry is broken and the adjoint fermions become massive and decouple, leaving behind the  anti-screening effect of the gauge fields. The theory confines deep in the IR.  If, however, $\alpha^*>\alpha_c$, then the theory avoids the triggering point and continues flowing to a conformal point. A third scenario is that the theory confines without chiral symmetry breaking. In this case, the theory has to have massless composite fermions that match various 't Hooft anomalies, see \cite{Anber:2018tcj,Cordova:2018acb,Bi:2018xvr,Wan:2018djl,Poppitz:2019fnp} for recent developments.  

\begin {table}
\begin{center}
\tabcolsep=0.11cm
\footnotesize
\begin{tabular}{|c|c|c|c|c|}
\hline
$n_G$ & $n_{\cal R}$ & $\alpha^*$ & $n_{\cal R}^*$ & Semi-classics without DTD?\\
\hline\hline
 $0$ & $[0,10]$ & $n_{\cal R}\in [5,10]\rightarrow \alpha^*\in [3.04,0.05]$ & $5$ & for $n_{\cal R}=\{0\}$ \\
 \hline
 $1$ & $[0,8]$ & $ n_{\cal R}\in[3,8] \rightarrow \alpha^*\in [2.60,0.05]$ & $6.42$  &for $n_{\cal R}=\{0\}$\\
 \hline
  $2$ & $[0,6]$ & $ n_{\cal R}\in[1,6] \rightarrow \alpha^*\in [2.23,0.05]$ & $4.38$ &for $n_{\cal R}=\{0\}$\\
  \hline
  $3$ & $[0,4]$ & $ n_{\cal R}\in[0,4] \rightarrow \alpha^*\in [0.75,0.05]$ & $2.34$ & for $n_{\cal R}=\{1\}$\\
  \hline
  $4$ & $[0,2]$ & $ n_{\cal R}\in[0,2] \rightarrow \alpha^*\in [0.21,0.05]$ & $0.31$ &  for $n_{\cal R}=\{1,2\}$\\
  \hline
  $5$ & $\{0\}$ & $n_{\cal R}=\{0\} \rightarrow \alpha^*=0.05$ &  &  for $n_{\cal R}=\{0\}$\\
  \hline
\end{tabular}
\caption{\label{classification} For a given number of adjoint flavors  $n_G$ (the first column)  we list the number of ${\cal R}$ flavors $n_{\cal R}$ (the second column) that guarantee that the theory stays asymptotically free. We also list the number of $\cal R$ flavors that allow the theory to have a fixed point (third column) and give the value range of the corresponding coupling constant $\alpha^*\equiv \frac{g_*^2}{4\pi}$  at the fixed point. In the fourth column we list the critical number of ${\cal R}$ fermions, above which the theory flows to a fixed point in the IR. For a later convenience, in the last column we list the number of $\cal R$ flavors that allow the theory to enter the semi-classical regime upon compactifying on a small circle without adding a double-trace deformation (DTD).}
\end{center}
\end{table}

One cannot directly apply the above logic to fermions in the $3$-index antisymmetric representation since the fermion bilinear in this representation vanishes identically: $\epsilon_{i_1i_2i_3j_1j_2j_3}$\newline$\times\epsilon_{\alpha\beta}\lambda^{\alpha\,i_1i_2i_3}\lambda^{\beta\,j_1j_2j_3}=0$.  Thus,  ${\cal R}$ fermions  will always contribute to the $\beta$-function (via the vacuum polarization graphs) as we flow from UV to IR. Nonetheless, one may still probe the breaking of chiral  symmetry via higher-dimensional operators, see, e.g., Eqs. (\ref{possible condensate}) and (\ref{Operator for chiSB}). Then, a theory with $n_{{\cal R}}$ fermions (and no adjoints) will either flow to a strongly coupled confining regime (with or without symmetry breaking), if the $\beta$-function does not develop a fixed point,  or to a CFT otherwise. Indeed, if the fixed point happens at a coupling const $\alpha^*\gtrsim 1$, then no robust conclusions can be reached. 

The story becomes more interesting when we have both adjoint and ${\cal R}$ fermions. Let us assume that as we flow from high to low scale we hit $\alpha^*$ before $\alpha_c=\frac{\pi}{18}$. Then, the theory flows to a CFT deep in the IR\footnote{We assume that a chiral phase transition for the ${\cal R}$ fermions does not set in before hitting  $\alpha^*$. If we apply (\ref{ladder approximations value}) naively by replacing $C_2(G)$ with $C_2({\cal R})$, we find that this is always the case. Of course, one cannot trust (\ref{ladder approximations value}) for fermions in the $3$-index antisymmetric representation since the fermion bilinear is zero. Finding the condition under which chiral phase transition breaks (via higher order operators) in this case is an interesting project that is left for the future.}. The condition $\alpha^*>\alpha_c$ implies that there is a lower bound on $n_{\cal R}$ for the theory to be inside the conformal window:
\begin{eqnarray}
n_{\cal R}>n_{\cal R}^*=\frac{2\left( 1992-480 n_G\right)}{471}\,,\quad \mbox{for}\, n_G\geq 1\,.
\end{eqnarray}
Using the two-loop $\beta$-function we find $\alpha\cong\alpha_c\cong 0.175$ at the critical value of $n_{\cal R}^*$, irrespective of the number of adjoint fermions. The smallness of the coupling constant indicates that it is very plausible that the theory will flow to a fixed point when the number of $\cal R$ fermions is bigger than $n_{\cal R}^*$.
For values of  $n_{\cal R}<n_{\cal R}^*$ we encounter $\alpha_c$ before $\alpha^*$ and both the global $U_{G-{\cal R}}(1)$ and adjoint chiral symmetries break. At energies below $\alpha_c$   the adjoints decouple leaving behind the gauge fields and ${\cal R}$ fermions to decide the fate of the $\beta$-function. For $n_{G}\geq 2$  and $n_{R}<n_{\cal R}^*$, we always find that at energy scales below $\alpha_c$ the theory flows to a regime with no fixed point deep in the IR: the theory confines with or without breaking its symmetries \footnote{Yet, a more intricate  scenario happens in the case of a single adjoint. In this case we have $n_{\cal R}^*=6.42$.  If we take $n_{\cal R}< 5$, then the theory does not develop a fixed point below  $\alpha_c$: it will presumably confine deep in the IR with or without breaking of the remaining symmetries. If we take $n_{\cal R}=\{5,6\}$, however,  then at energy scales below   $\alpha_c$  the theory develops a fixed point at  $\alpha^*=\{3.04,0.891\}$. This is puzzling and we expect that the $2$-loop $\beta$-function might not be robust for such conclusion.}.

 The above discussion was a general view on the phase structure of the theory on $\mathbb R^4$. In the next two sections we closely examine the self-conjugate QCD: first by analyzing the theory on $\mathbb R^3\times \mathbb S^1_L$ and then by examining the various 't Hooft anomaly matching conditions. Using this analysis, we will be able to draw more concrete conclusions about the possible IR spectrum of the theory on $\mathbb R^4$ and whether or not this spectrum is adiabatically connected to the theory on the circle.

\section{Compactifying on $\mathbb S^1_L$: the semi-classical analysis, (non)perturbative spectrum, and  mass gap}
\label{Semi-classical treatment}

In this section we carry out a systematic study of the theory\footnote{See \cite{Poppitz:2009kz} for a similar story on $\mathbb R^3\times \mathbb S^1_L$  for a supersymmetric theory without a bilinear mass term.} on $\mathbb R^3\times \mathbb S_L^1$. Our starting point is the potential (\ref{potential in terms of the polyakov loop}) to determine whether it is minimized inside the affine Weyl chamber\footnote{\label{Weyl chamber note}The affine Weyl chamber is the region of physically inequivalent values of $\bm \Phi$. For $SU(N)$, this region is a polyhedron in $N-1$ dimensional space defined by the inequalities
\begin{eqnarray}
\bm \alpha_a\cdot\bm \Phi>0 \quad \mbox{for all}\quad a=1,2,3,...,N-1\quad \mbox{and} \quad-\bm\alpha_0\cdot \bm \Phi<2\pi\,.
\end{eqnarray}
The interior of this region (not including the boundary) is the smallest region in the $\bm \Phi$-space with no massless $W$-bosons (including their Kaluza-Klein excitations), see \cite{Argyres:2012ka,Anber:2017pak} for more details.} of $\bm \Phi$. If this is the case, then the Kaluza-Klein excitations are heavy (the masses are $\frac{2n\pi }{6L}$ and $n=1,2,....$) and the $4$-dimensional coupling constant $g$ ceases to run at scale $\sim \frac{1}{L}$. Thus, by taking $L$ to be small enough such that $L\Lambda \ll1$, where $\Lambda$ is the strong scale, we can guarantee that our theory is weakly coupled and amenable to the semi-classical analysis. A detailed study of the potential $V\left(\bm \Phi\right)$ was carried out in \cite{Anber:2017pak}  and it was concluded that the potential is minimized inside the affine Weyl chamber for the following number of $G$ and ${\cal R}$ flavors: 
\begin{eqnarray}
\left(n_{G}, n_{{\cal R}}\right)=(3,1)\,, (4,1)\,, (4,2)\,.
\label{values for stabilization}
\end{eqnarray}  
Thus, a small value of $n_G$ is not enough to compete against the gauge and ${\cal R}$ fermion fluctuations, which push the minimum to the boundary of the affine Weyl Chamber, see Footnote \ref{Weyl chamber note}. Also, higher values of $n_G$  will push the theory outside its asymptotic freedom window. For the values of $n_G$ and $n_{\cal R}$ quoted in (\ref{values for stabilization}), it was shown in \cite{Anber:2017pak} that the global minimum of $V\left(\bm \Phi\right)$ is a center-symmetric point:
\begin{eqnarray}
\bm\Phi_0=\frac{2\pi}{6}\bm \rho\,,
\end{eqnarray}
where $\bm\rho$ is the Weyl vector given by $\bm \rho=\sum_{a=1}^{5}\bm \omega_a$ and $\bm\omega_{a}$, $a=1,2,...,5$ are the fundamental weights.

For small values of $L$ we can dimensionally reduce the theory from $4$ to $3$ dimensions.  At the center symmetric point the gauge group $SU(6)$ spontaneously breaks into the maximal abelian subgroup $U(1)^{5}$. Then, it can  be shown that the adjoint fermions in the Cartan directions, $\bm \chi$, are neutral under $U(1)^{5}$ and stay massless, while  the $\lambda$ fermion components that are charged under $U(1)^{5}$ are massive\footnote{It is important to emphasize that here we are referring to  the $3$-D mass term. It results from the dimensional reduction of the kinetic term from $4$- to $3$-D, and hence, it is nonvanishing, unlike the $4$-D mass term.}, preventing the theory from flowing to a strongly coupled point in the IR, see  \cite{Anber:2017pak} for details. Further, we can use the $3$-D  abelian duality $\bm F_{\mu\nu}\sim\epsilon_{\mu\nu\alpha}\partial_\alpha\bm \sigma$, where $\bm \sigma$ is a compact scalar (the dual photon), to write the perturbative $3$-D effective Lagrangian:
\begin{eqnarray}
{\cal L}_{3-D,\mbox{pert}}=\left(\partial_\mu\bm \sigma\right)^2+i\bar{\bm \chi}_p\bar\sigma^\mu\partial_\mu \bm\chi^p\,,
\label{3D perturbative Lagrangian}
\end{eqnarray}
where $\mu=0,1,2$. This concludes the perturbative analysis of the theory on $\mathbb R^3\times \mathbb S^1_L$.

\subsection*{Double-trace deformation}

As we discussed above, one needs at least $3$ adjoint fermions in order for the global minimum of $V(\bm \Phi)$ to lie inside the affine Weyl chamber. This was necessary for the breaking of $SU(N)$ to its maximal abelian subgroup, and hence, being able to use semi-classical analytical techniques to analyze the theory. For a lower number of $n_G$, however, one can still obtain the desired breaking by adding a double-trace deformation to the original Lagrangian (\ref{Lagrangian of the system}). This takes the form
\begin{eqnarray}
\Delta {\cal L}_{DTD}=\frac{1}{L^3}\sum_{n=1}a_n |\mbox{tr}_F\left[\Omega^n\right]|^2\,,
\end{eqnarray}
with sufficiently positive coefficients $a_n$ that guarantee that  $V(\bm \Phi)$ will attain a global minimum at the center-symmetric point $\bm \Phi_0$. Repeating the arguments of the previous paragraph, we arrive to the same effective $3$-D Lagrangian given by (\ref{3D perturbative Lagrangian}).

\subsection*{The index theorem and monopole operators}

In addition to the perturbative spectrum, the theory admits nonperturbative saddles in the form of monopole instantons. The number of the fermion zero modes in the background of these monopoles is given by the index theorem \cite{Atiyah:1968mp,Nye:2000eg,Poppitz:2008hr}. Let the action\footnote{Since we are working at the center-symmetric point $\bm \Phi_0$, all the fundamental monopoles have the same action irrespective of their charges.} of a fundamental monopole\footnote{A $SU(N)$ BPST instanton with a unit topological charge is made of $N$ fundamental monopole instantons, each contributes a  fraction $\frac{1}{N}$ of the topological charge. See \cite{Kraan:1998pm,Anber:2011de} for details.} charged under $\bm \alpha_a$, $a=0,1,..,5$, be $S_m$, where $S_m=\frac{8\pi^2}{6g^2}$ at the center-symmetric point $\bm\Phi_0$.  Then, there are $2$ adjoint fermion zero modes in the background of each fundamental monopole. The situation in the case of $\cal R$ fermions is more involved. For a later convenience,  we give the number of fermion zero modes, $n_f$, in the background of a monopole with a general action $S=nS_m$:
\begin{eqnarray}
\nonumber
&&\mbox{for}\,\, n \in 2\mathbb Z\,, \quad\quad\quad n_f=n\,\, \mbox{for all monopoles with charges}\,\, \bm \alpha_{0,1,2,3,4,5}\,,\\
\nonumber
&&\mbox{for}\,\, n \in 2\mathbb Z+1\,, \quad n_f=n-1\,\, \mbox{ for monopoles with charges}\,\, \bm \alpha_{0,2,4}\,,\\
&&\quad \quad \quad\quad\quad\quad\quad\quad n_f=n+1\,\, \mbox{ for monopoles with charges}\,\, \bm \alpha_{1,3,5}\,.
\label{index theorem and fermion zero modes}
\end{eqnarray}
The index theorem dictates the form of the monopole operators. The schematic form of the fundamental-monopole operators reads (here we assume $n_G>0$ and $n_{\cal R}>1$ leaving the special cases $n_G=0$ and $n_{\cal R}=1$ for separate sections):
\begin{eqnarray}
\nonumber
{\cal M}_{0}&=&e^{-S_m}e^{i\bm \alpha_0\cdot \bm \sigma}\left(\chi\chi\right)^{n_G}\,,\quad {\cal M}_{1}=e^{-S_m}e^{i\bm \alpha_1\cdot \bm \sigma}\left(\chi\chi\right)^{n_G}\left(\lambda\lambda\right)^{n_{\cal R}}\,,\\
\nonumber
{\cal M}_{2}&=&e^{-S_m}e^{i\bm \alpha_2\cdot \bm \sigma}\left(\chi\chi\right)^{n_G}\,,\quad {\cal M}_{3}=e^{-S_m}e^{i\bm \alpha_3\cdot \bm \sigma}\left(\chi\chi\right)^{n_G}\left(\lambda\lambda\right)^{n_{\cal R}}\,,\\
{\cal M}_{4}&=&e^{-S_m}e^{i\bm \alpha_4\cdot \bm \sigma}\left(\chi\chi\right)^{n_G}\,,\quad{\cal M}_{5}=e^{-S_m}e^{i\bm \alpha_5\cdot \bm \sigma}\left(\chi\chi\right)^{n_G}\left(\lambda\lambda\right)^{n_{\cal R}}\,.
\label{monopole vortices}
\end{eqnarray} 
The anti-monopole operator $\overline {{\cal M}_a}$ can be obtained by the assignment $\alpha_a\rightarrow -\alpha_a$, $\chi\rightarrow \bar \chi$, and $\lambda\rightarrow \bar\lambda$.

The monopole operators are also invariant under the $0$-form symmetries in (\ref{all good symmetries}). The invariance under $SU(n_G)$ and $SU(n_{\cal R})$ is evident. The invariance under the abelian group $\mathbb Z^{G}_{12n_G}\times \mathbb Z^{\cal R}_{6n_{\cal R}}\times U_{G-{\cal R}}(1)$ forces the dual photon $\bm \sigma$ to transform under these symmetries. In other words, the abelian symmetry action on fermions is intertwined with  shift symmetries acting  on $\bm \sigma$. 

Now, we comment on the fate of the $1$-form center symmetry $\mathbb Z_3^{(1)}$ in $3$-D. Upon dimensional reduction from $4$- to $3$-D, the $1$-form center symmetry that wraps $\mathbb S^1_L$   becomes a $0$-form $\mathbb Z_3^{(0)}$ center symmetry from the point of view of observers in $3$-D. The actions of this symmetry on  $\bm \sigma$ is given by $\bm \sigma \rightarrow {\cal P}\bm\sigma$, where ${\cal P}$ is  a $\mathbb Z_3$ cyclic permutation operator, see \cite{Anber:2015wha}. Therefore, $\mathbb Z_3^{(0)}$ permutes the operators $\{{\cal M}_{0},{\cal M}_{2},{\cal M}_{4}\}$. Similarly, it permutes the operators $\{{\cal M}_{1},{\cal M}_{3},{\cal M}_{5}\}$. In other words, $\mathbb Z_3^{(0)}$ preserves the structure of the fermion zero modes in the monopole background. A summary of the abelian symmetries, the charges of the fermions, and the action of the  symmetries on the dual photon is provided in Table \ref{symmetries}.

\begin {table}
\begin{center}
\tabcolsep=0.11cm
\footnotesize
\begin{tabular}{|c|c|c|c|}
\hline
 Symmetry & $\chi$ & $\lambda$ & Action on $\bm \sigma$\\
 \hline\hline
 $U_{G-{\cal R}}(1)$ & $n_{\cal R}\alpha$ & $-2n_G\alpha$ &$\bm \sigma \rightarrow \bm\sigma+n_Gn_{\cal R}\alpha\left(\bm \omega_1-\bm \omega_2+\bm\omega_3-\bm \omega_4+\bm \omega_5\right)$\\
\hline
$\mathbb Z_{6n_{\cal R}}^{{\cal R}}$ & $0$ & $1$ & $\bm \sigma\rightarrow \bm\sigma -\frac{4\pi}{6}\left(\bm \omega_1+\bm\omega_3+\bm\omega_5\right)$\\
\hline
$\mathbb Z_{12n_G}^G$ & $1$  & $0$ & $\bm\sigma\rightarrow \bm \sigma -\frac{2\pi}{6}\bm\rho$\\\hline
$\mathbb Z_{3}^{(0)}$ & $0$ & $0$ & $\bm\sigma\rightarrow{\cal P}\bm\sigma$\\
\hline
\end{tabular}
\caption{\label{symmetries}A summary of the abelian symmetries,  the fermion charges, and the action of the  symmetries on the dual photon.}
\end{center}
\end{table}

\subsection*{Bions and the mass gap  in the general case $n_G> 0$ and $n_{\cal R}>1$}

The monopole instantons cannot give rise to a mass gap since they are dressed with fermion zero modes. Nonetheless, complex molecules made of two monopoles can form, proliferate in the vacuum, and generate a mass gap. This can happen if the fermion zero modes are soaked up inside the molecules. Such structures were first identified in \cite{Unsal:2007jx} and were dubbed the {\em magentic bions}. The bions in the theory at hand are:
\begin{eqnarray}
\nonumber
{\cal B}_1&=&{\cal M}_0\overline {\cal M}_2=f_1(g)e^{-2S_m}e^{i(\bm \alpha_0-\bm \alpha_2)\cdot \bm \sigma}\,,\quad
{\cal B}_4={\cal M}_1\overline {\cal M}_3=f_2(g)e^{-2S_m}e^{i(\bm \alpha_1-\bm \alpha_3)\cdot \bm \sigma}\,,\\
\nonumber
{\cal B}_2&=&{\cal M}_2\overline {\cal M}_4=f_1(g)e^{-2S_m}e^{i(\bm \alpha_2-\bm \alpha_4)\cdot \bm \sigma}\,,\quad
{\cal B}_5={\cal M}_3\overline {\cal M}_5=f_2(g)e^{-2S_m}e^{i(\bm \alpha_3-\bm \alpha_5)\cdot \bm \sigma}\,,\\
{\cal B}_3&=&{\cal M}_4\overline {\cal M}_0=f_1(g)e^{-2S_m}e^{i(\bm \alpha_4-\bm \alpha_0)\cdot \bm \sigma}\,,\quad
{\cal B}_6={\cal M}_5\overline {\cal M}_1=f_2(g)e^{-2S_m}e^{i(\bm \alpha_5-\bm \alpha_1)\cdot \bm \sigma}\,,
\end{eqnarray}
where $f_1(g)$ and $f_2(g)$ are prefactors that will not play a significant role in our analysis\footnote{The functional form of $f_1(g)$ and $f_2(g)$ can be determined by counting the number of the moduli and the fermion zero modes in the background of bions. Following the analysis in \cite{Anber:2011de} we find:
\begin{eqnarray}
f_1(g)=\frac{1}{g^{8-8n_G}}\,,\quad f_2(g)=\frac{1}{g^{8-8(n_G+n_{\cal R})}}\,.
\end{eqnarray}
}. Notice that the action of the $0$-form center symmetry $\mathbb Z_3^{(0)}$ on the bions is  ${\cal B}_1\rightarrow {\cal B}_3 \rightarrow {\cal B}_5$ and ${\cal B}_2\rightarrow {\cal B}_4\rightarrow {\cal B}_6$, and hence, it preserves the form of $f_1(g)$ and $f_2(g)$. In addition, all the bions are invariant under the symmetries in  (\ref{all good symmetries}).

The effect of the bions can be taken into account in the partition function by summing over their paths. Now, recalling the perturbative $3$-D effective Lagrangian (\ref{3D perturbative Lagrangian}) and adding the contribution from the monopoles and bions we finally obtain (working at a center-symmetric point, with or without a double-trace deformation):
\begin{eqnarray}
{\cal L}_{3-D,\mbox{total}}=\left(\partial_\mu\bm \sigma\right)^2+V(\bm\sigma)+i\bar{\bm \chi}_p\bar\sigma^\mu\partial_\mu \bm\chi^p+\sum_{i=1}^6\left({\cal M}_i+\overline{\cal M}_i\right)\,,
\label{3D perturbative Lagrangian with potential}
\end{eqnarray}
where
\begin{eqnarray}
\nonumber
L^2V(\bm \sigma)&=&\sum_{i=1}^{6}\left({\cal B}_i+\overline{\cal B}_i\right)\\
\nonumber
&=&-e^{-2S_m}f_1(g)\left\{\cos\left[\left(\bm\alpha_{0}-\bm \alpha_2\right)\cdot \bm\sigma\right] +\cos\left[\left(\bm\alpha_{2}-\bm \alpha_4\right)\cdot \bm\sigma\right] +\cos\left[\left(\bm\alpha_{4}-\bm \alpha_0\right)\cdot \bm\sigma\right] \right\}\\
\nonumber
&&-e^{-2S_m}f_2(g)\left\{\cos\left[\left(\bm\alpha_{1}-\bm \alpha_3\right)\cdot \bm\sigma\right] +\cos\left[\left(\bm\alpha_{3}-\bm \alpha_5\right)\cdot \bm\sigma\right] +\cos\left[\left(\bm\alpha_{5}-\bm \alpha_1\right)\cdot \bm\sigma\right] \right\}\,.\\
\label{photon potential}
\end{eqnarray}

In the next sections we examine the vacuum of the theory for various numbers of fermions in the adjoint and $3$-index antisymmetric representations. 

\subsection*{Vacuum structure in the general case  $n_G > 0$ and $n_{\cal R}>1$ }

First, we need to determine the fundamental domain of the dual photon. We recall that the gauge group $SU(6)$ is the covering group, and therefore, the smallest allowed electric probes are valued in the weight lattice. The monodromy  $\oint_C d\bm\sigma$  is proportional to the  electric charge (flux) enclosed by $C$. Then, it immediately follows that the fundamental domain of the dual photon is $\bm\sigma\equiv \bm\sigma+2\pi\bm\omega_a$ for all $a=1,2,...,5$. 

In order to find the global minima of the potential (\ref{photon potential}) we assume that the solution is valued in the weight lattice, i.e., we write
\begin{eqnarray}
\bm \sigma_0=\sum_{a=1}^5 C_a\bm\omega_a\,,
\label{general solution adj}
\end{eqnarray}
for arbitrary coefficients $C_a$ to be determined momentarily. Furthermore, we express the roots and weights in the $\mathbb R^N$ basis \footnote{Recall that the dimension of the Cartan subspace of $SU(N)$ is $N-1$, which is also the number of the dual photons. In the $\mathbb R^N$ basis we add one extra dimension, and hence, one spurious degree of freedom. In order to eliminate  the unphysical mode, we impose the condition $\sum_{i=1}^N\sigma_i=0$. See \cite{Argyres:2012ka} for details.}. Then, it is a simple exercise to show that the global minima of $V(\bm \sigma)$ in the fundamental domain are given by
\begin{eqnarray}
\bm \sigma_0(k)=-\left(C_2+\frac{2\pi k}{3}\right)\left(\bm \omega_1+\bm\omega_3+\bm\omega_5\right)+C_2\left(\bm \omega_2+\bm\omega_4\right)\,,
\label{general solution for adj and 3-index}
\end{eqnarray}
for  $k=0,1,2$ and $C_2 \in \mathbb R$. This solution represents $3$ flat valleys separated by hills. The integer $k=0,1,2$ labels the valleys, while  $C_2$ parametrizes the flat direction along the valley.  The fact that there is a single modulus $C_2$ is consistent with the existence of a single global $U_{G-{\cal R}}(1)$ symmetry. 

The dual photon masses are given by 
\begin{eqnarray}
\nonumber
M^2_{\scriptsize\mbox{photon}}&=&\frac{1}{2}\left[\frac{\partial ^2 V(\bm\sigma)}{\partial\sigma_i\partial\sigma_j}\right]_{\bm\sigma_{0}(k)}=\\
&&\frac{3}{2L^2}e^{-2S_m}\mbox{diag}\left(f_1+f_2+f,f_1+f_2+f,f_1+f_2-f,f_1+f_2-f,0\right)\,,
\end{eqnarray}
where $f\equiv\sqrt{f_1^2-f_1f_2+f_2^2}$. As expected, there is a single massless photon, which is in accord with the existence of a single $U_{G-\cal R}(1)$ global symmetry. 

As we discussed above, the adjoint fermion components along the Cartan subspace are massless in the perturbative sector. These transform in the fundamental representation of $SU(n_G)$.  Adding the monopoles does not change this picture, except for the special case $n_G=1$. Here, the monopole operators  generate a fermion mass term when evaluated in the ground state:
\begin{eqnarray}
\left\langle \sum_{i=1}^6 {\cal M}_i+\bar{\cal M}_i \right\rangle(\bm\sigma_0)=\frac{3}{L}e^{-S_m}\left[e^{iC_2}\bm\chi\bm\chi+\mbox{h.c.} \right]\,.
\end{eqnarray} 
Therefore, a theory with a single adjoint fermion has only a single bosonic (photon) massless excitation deep in the IR.  One can  think of the massless photon in this case as the Goldstone boson that results from breaking $U_{G-{\cal R}}(1)$ chiral symmetry; this  breaking generates a mass term for the adjoint fermion\footnote{We would like to thank Erich Poppitz for emphasizing this point.}.

\subsection*{The special case $n_G=0$ and $n_{\cal R}>1$}
The absence of adjoint fermions means that the single classical $U_{\cal R}(1)$ symmetry  becomes anomalous on the quantum level. Then, the global symmetry of the theory is
\begin{eqnarray}
\frac{SU(n_{\cal R})\times \mathbb Z_{6n_{\cal R}}^{\cal R}}{\mathbb Z_{n_{\cal R}}}\times \mathbb Z_3^{(1)}\,.
\end{eqnarray}
The absence of adjoint fermions demands that we add a double trace deformation to bring the theory to its semi-classical regime. 
In this special case the monopoles charged under $\bm \alpha_0, \bm \alpha_2, \bm\alpha_3$ do not carry fermion zero modes. These monopoles  proliferate in the vacuum and participate in the generation of the mass gap.  Then, the dual photon potential receives contributions from both  the monopoles ${\cal M}_0,{\cal M}_2,{\cal M}_4$ and the bions ${\cal B}_4,{\cal B}_5,{\cal B}_5$. The total potential reads:
\begin{eqnarray}
\nonumber
L^2V(\bm\sigma)&=&-f_3(g)e^{-S_m}\left(\cos\left(\bm\sigma\cdot\bm\alpha_0\right)+\cos\left(\bm\sigma\cdot\bm\alpha_2\right)+\cos\left(\bm\sigma\cdot\bm\alpha_4\right)\right)\\
\nonumber
&&-f_4(g)e^{-2S_m}\left(\cos\left(\bm\sigma\cdot\left(\bm\alpha_1-\bm\alpha_3\right)\right)+\cos\left(\bm\sigma\cdot\left(\bm\alpha_3-\bm\alpha_5\right)\right)+\cos\left(\bm\sigma\cdot\left(\bm\alpha_5-\bm\alpha_1\right)\right)\right)\,.\\
\label{potential for the first special case}
\end{eqnarray}
The global minima of this potential inside the fundamental domain of $\bm\sigma$ are
\begin{eqnarray}
\bm\sigma_0(k)=\frac{2\pi k }{3}(\bm \omega_1+\bm \omega_3+\bm \omega_5)\,,
\label{global minima for the first case}
\end{eqnarray}
for $k=0,1,2$. The photon masses are given by (neglecting the prefactors $f_3,f_4$ and using the fact $e^{-2S_m}\ll e^{-S_m}$):
\begin{eqnarray}
M_{\scriptsize\mbox{photon}}^2\cong \frac{1}{L^2}e^{-S_m}\mbox{diag}\left(1,1+\frac{3}{4}e^{-S_m},1+\frac{3}{4}e^{-S_m},\frac{9}{4}e^{-S_m},\frac{9}{4}e^{-S_m}\right)\,.
\label{mass gap for nRG1 and nG0}
\end{eqnarray}
We immediately see that all the photons are massive, which sounds with the fact that the special case at hand does not enjoy a global $U(1)$ symmetry. However, not all of these masses are of the same order; $2$ out of the $5$ photons have exponentially small masses compared to the rest. This can be understood easily from the structure of the potential (\ref{potential for the first special case}). The first term (the monopole contribution) gives mass to $3$ photons leaving $2$ flat directions. The latter get lifted by the second term in (\ref{potential for the first special case}), which is the next-to-leading order contribution to the mass gap (the bion contribution). We conclude that this theory is fully gapped in the IR.

\subsection*{The special case $n_G=0$ and $n_{\cal R}=1$}

Here, the global symmetry is simply $\mathbb Z_6^{\cal R}\times \mathbb Z_3^{(1)}$. Again, the monopoles charged under $\bm \alpha_0, \bm \alpha_2, \bm\alpha_3$ do not carry fermion zero modes  and generate masses to $3$ of out of the $5$ photons. However, unlike the previous cases, the monopole operators  ${\cal M}_{1},{\cal M}_{2},{\cal M}_{3}$ vanish identically because of the vanishing of the fermion bilinear $\lambda\lambda$. More specifically, because $\lambda$ is in the $3$-index antisymmetric representation we have
\begin{eqnarray}
\epsilon_{i_1i_2i_3i_4i_5i_6}\epsilon_{\alpha\beta}\lambda^{\alpha\,i_1i_2i_3}\lambda^{\beta\,i_4i_5i_6}=0\,,
\end{eqnarray}
where $i_1,i_2,...,i_6$ are the color indices of the $3$-index anti-symmetric representation and $\alpha,\beta$ are the indices of the Lorentz group. The vanishing of the fermion bilinear is attributed to having two Levi-Civita tensors along with the Grassmannian nature of fermions. Using the result  (\ref{index theorem and fermion zero modes}) from the index theorem, we realize that the first nonvanishing monopole operator has $3$ times the action of a fundamental monopole $S=3S_m$, which carries $4$ fermion zero modes. These are the monopoles charged under $\bm\alpha_1, \bm\alpha_3,\bm\alpha_5$. These higher-order monopoles form higher-order bions of the form ${\cal M}_1^{(3)}\overline{{\cal M}_3^{(3)}}\sim e^{-6S_m}e^{i\bm\sigma \left(\bm\alpha_1-\bm\alpha_3\right)}$, etc. Taking the contribution from all the saddles we arrive to the total potential:
\begin{eqnarray}
\nonumber
L^2V(\bm\sigma)&=&-f_5(g)e^{-S_m}\left(\cos\left(\bm\sigma\cdot\bm\alpha_0\right)+\cos\left(\bm\sigma\cdot\bm\alpha_2\right)+\cos\left(\bm\sigma\cdot\bm\alpha_4\right)\right)\\
\nonumber
&&-f_6(g)e^{-6S_m}\left(\cos\left(\bm\sigma\cdot\left(\bm\alpha_1-\bm\alpha_3\right)\right)+\cos\left(\bm\sigma\cdot\left(\bm\alpha_3-\bm\alpha_5\right)\right)+\cos\left(\bm\sigma\cdot\left(\bm\alpha_5-\bm\alpha_1\right)\right)\right)\,.\\
\label{potential for the second special case}
\end{eqnarray}
This potential has global minima given by (\ref{global minima for the first case}), where all the photons acquire masses: 
\begin{eqnarray}
M_{\scriptsize\mbox{photon}}^2\cong \frac{1}{L^2}e^{-S_m}\mbox{diag}\left(1,1+\frac{3}{4}e^{-5S_m},1+\frac{3}{4}e^{-5S_m},\frac{9}{4}e^{-5S_m},\frac{9}{4}e^{-5S_m}\right)\,.
\label{photon mass for nR1 and nG0}
\end{eqnarray}
Again, this theory is fully gapped in the IR. 

\subsection*{The special case $n_G>0$ and $n_{\cal R}=1$}
The global symmetry of the theory is
\begin{eqnarray}
\frac{SU(n_G)\times \mathbb Z^{G}_{12n_G}}{\mathbb Z_{n_G}}\times \mathbb Z^{\cal R}_6\times U_{G-{\cal R}}(1)\times\mathbb Z^{(1)}_3\,.
\label{all good symmetries for the last case}
\end{eqnarray}
The mass gap receives contribution from the bions made of the fundamental monopoles charged under $\bm\alpha_0,\bm\alpha_2,\bm\alpha_4$ and  bions made of the third-order monopoles charged under $\bm\alpha_1,\bm\alpha_3,\bm\alpha_5$. The total potential reads
\begin{eqnarray}
\nonumber
L^2V(\bm \sigma)&=&-e^{-2S_m}f_7(g)\left\{\cos\left[\left(\bm\alpha_{0}-\bm \alpha_2\right)\cdot \bm\sigma\right] +\cos\left[\left(\bm\alpha_{2}-\bm \alpha_4\right)\cdot \bm\sigma\right] +\cos\left[\left(\bm\alpha_{4}-\bm \alpha_0\right)\cdot \bm\sigma\right] \right\}\\
\nonumber
&&-e^{-6S_m}f_8(g)\left\{\cos\left[\left(\bm\alpha_{1}-\bm \alpha_3\right)\cdot \bm\sigma\right] +\cos\left[\left(\bm\alpha_{3}-\bm \alpha_5\right)\cdot \bm\sigma\right] +\cos\left[\left(\bm\alpha_{5}-\bm \alpha_1\right)\cdot \bm\sigma\right] \right\}\,.\\
\label{photon potential for the last case}
\end{eqnarray}
The global minima of this potential are given by (\ref{general solution for adj and 3-index}), which also reflects the invariance of the solutions under $U_{G-\cal R}(1)$ global symmetry. Now, one of the $5$ photons remains massless, while $2$ photons acquire a mass  $\sim e^{-S_m}/L$ and the remaining $2$ photons acquire a mass  $\sim e^{-3S_m}/L$.

\subsection*{The IR particle spectrum on $\mathbb R^3\times \mathbb S^1_L$:  a summary}

We end this section by summarizing the spectrum of the theory deep in the IR. The theory on $\mathbb R^3\times \mathbb S_L^1$ lives at a center-symmetric point, i.e., the potential $V(\bm\Phi)$ has a global minimum at $\bm \Phi_0=\frac{2\pi}{6}\bm \rho$. This can be achieved either by adding  adjoint fermions, $n_G\geq 3$, or by supplementing the theory with a double-trace deformation. In all cases we found that the theory admits $3$ degenerate ground states within the fundamental domain of the dual photon. This implies the following breaking pattern of the  discrete symmetries:
\begin{eqnarray}
\mathbb Z^{\cal R}_{6 n_{\cal R}}\rightarrow \mathbb Z_{2n_{\cal R}}\,,\quad \mathbb Z_{12 n_G}^{G}\rightarrow \mathbb Z_{4n_G}\,.
\label{the breaking pattern}
\end{eqnarray}
As we show in the next section, this degeneracy and symmetry breaking pattern is  the consequence of a mixed 't Hooft anomaly between the $0$-form discrete symmetries and the $1$-form $\mathbb Z_3^{(1)}$ center symmetry. In the absence of adjoint fermions the spectrum is fully gapped in the IR.  Adding adjoint fermions endows the theory with a $U_{G-{\cal R}}(1)$ global symmetry. As a result, one of the photons, $\sigma$, stays massless. The components of the adjoint fermions along the Cartan subspace are also massless, except when $n_G=1$. In this case the single adjoint fermion acquires a mass via the monopole operators. Thus, the IR spectrum of the theory with adjoints (and at least $1$ fermion in $\cal R$) is summarized as:
\begin{eqnarray}
{\cal L}_{\scriptsize {IR}}=\left(\partial_\mu\sigma\right)^2+\epsilon\sum_{p=1}^{n_G}\bar{\bm\chi_p}\partial_\mu\bar\sigma^\mu\bm\chi^p\,,
\end{eqnarray}
where   $\epsilon=0$ only and only if $n_G=1$ and $1$ otherwise. In all cases, with or without adjoints, the $1$-form $\mathbb Z^{(1)}_3$ center symmetry remains intact:  the theory develops a mass gap and confines the fundamental charges. These cases are also summarized in Table \ref{all cases summary} .

The breaking of the discrete chiral symmetries lead to domain walls that interpolate between the degenerate vacua.   We expect that the strings between fundamental static quarks are composed of two domain walls,  as was shown in \cite{Anber:2015kea} for QCD(adj).

\begin {table}
\begin{center}
\tabcolsep=0.11cm
\footnotesize
\begin{tabular}{|l|l|l|}
\hline
\backslashbox{$n_{\cal R}$}{$n_G$} & $=0$ &  $>0$\\
\hline
$=1$ & $\diamond$  $3$ photons with mass $\sim\frac{1}{L}e^{-S_m/2}$ & $\diamond$  $2$ photons with mass $\sim\frac{1}{L}e^{-S_m}$,\\ 
 &  \hspace{1.5mm}   and $2$ photons with mass $\sim\frac{1}{L}e^{-3S_m}$ & \hspace{1.5mm} $2$ photons with mass $\sim\frac{1}{L}e^{-3S_m}$, and $1$ massless photon  \\
 & $\diamond$ vacua: $\bm \sigma_0(k)=\frac{2\pi k}{3}\left(\bm \omega_1+\bm\omega_3+\bm\omega_5\right)$  & $\diamond$ single massless $SU(n_G)$ fermion \\
  &  & \hspace{3mm}except when $n_G=1$, then the fermion gets a mass $\sim\frac{1}{L}e^{-S_m}$\\
 & & $\diamond$ vacua: $\bm \sigma_0(k)=-\left(C_2+\frac{2\pi k}{3}\right)\left(\bm \omega_1+\bm\omega_3+\bm\omega_5\right)+C_2\left(\bm \omega_2+\bm\omega_4\right)$ \\
\hline
$>1$ & $\diamond$  $3$ photons with mass $\sim\frac{1}{L}e^{-S_m/2}$ & $\diamond$  $4$ photons with mass $\sim\frac{1}{L}e^{-S_m}$ and $1$ massless photon  \\
 & \hspace{1.5mm}  and $2$ photons with mass $\sim\frac{1}{L}e^{-S_m}$    & $\diamond$ single massless $SU(n_G)$ fermion \\
  & $\diamond$ vacua: $\bm \sigma_0(k)=\frac{2\pi k}{3}\left(\bm \omega_1+\bm\omega_3+\bm\omega_5\right)$ &\hspace{1.5mm} except when $n_G=1$, then the fermion gets a mass $\sim\frac{1}{L}e^{-S_m}$\\
 & & $\diamond$ vacua: $\bm \sigma_0(k)=-\left(C_2+\frac{2\pi k}{3}\right)\left(\bm \omega_1+\bm\omega_3+\bm\omega_5\right)+C_2\left(\bm \omega_2+\bm\omega_4\right)$ \\
\hline
\end{tabular}
\caption{\label{all cases summary} A summary of the different cases on $\mathbb R^3\times \mathbb S^1_L$.}
\end{center}
\end{table}

\section{'t Hooft anomalies, the IR spectrum on $\mathbb R^4$, and adiabatic continuity}
\label{t Hooft anomalies and IR spectrum}

\subsection*{The $1$-form/$0$-form mixed anomaly}

't Hooft anomalies are obstructions to gauging global symmetries. They provide a unique handle to study the nonperturbative aspects of a theory and put sever constraints on its IR spectrum.
  
In this section we compute 't Hooft anomalies  and use them to track the theory as we decompactify $\mathbb S^1_L$. Of particular importance is the recently discovered mixed anomaly between the $0$-form discrete chiral symmetries and  $1$-form $\mathbb Z_3^{(1)}$ center symmetry \cite{Gaiotto:2017yup}. One way to see this anomaly in the UV is to gauge the center symmetry by turning on 't Hooft twists \cite{tHooft:1979rtg,Anber:2018xek,Anber:2018jdf}. In order to demonstaret how this works, we start by a warm up exercise and confirm that the theory is indeed invariant under the discrete symmetries in (\ref{all good symmetries}). Under global  $U_G(1)$ and $U_{\cal R}(1)$ rotations the fermions and measure transform as
\begin{eqnarray}
\nonumber
&&\chi\rightarrow e^{i\alpha_G}\chi\,,\lambda\rightarrow e^{i\alpha_{\cal R}}\lambda \implies[{\cal D}\chi][{\cal D}\lambda]\rightarrow e^{i\left(n_G\alpha_G T(G)+n_{\cal R}\alpha_{\cal R}T(\cal R)\right)Q_T} [D\chi][D\lambda]\,,
\end{eqnarray}
where $Q_T=\frac{1}{16\pi^2}\int_{{\cal M}^4} \mbox{tr}_F\left[F_{MN}\tilde F_{MN}\right]$ is the topological charge. Thus, we find that the measure is invariant under the following transformations:
\begin{eqnarray} 
\nonumber
\alpha_{{\cal R}}&=&-\frac{2n_G \alpha_G}{n_{{\cal R}}}\implies U_{G-{\cal R}}(1)\,,\\
\nonumber
\alpha_G&=&\frac{2\pi }{12n_G}\,, \alpha_{{\cal R}}=0  \implies \mathbb Z^{G}_{12n_G}\,,\quad
\nonumber
\alpha_{{\cal R}}=\frac{2\pi }{6n_{\cal R}}\,,\alpha_G=0 \implies \mathbb Z^{{\cal R}}_{6n_{\cal R}}\,.
\end{eqnarray}
Now we gauge the $1$-form symmetry:  we compactify $\mathbb R^4$ on a $4$-torus $\mathbb T^4$ and then turn on 't Hooft twists. We need to take  large enough $\mathbb T^4$  (the cycles should be larger than $\Lambda^{-1}$) in order to avoid phase transitions. Then, one can show that  the topological charge  is fractional  on $\mathbb T^4$:  $Q_T=\frac{k}{3}$, for $k={0,1,2}$. Finally,  under a discrete symmetry transformation ($Z^{G}_{12n_G}$ or $Z^{{\cal R}}_{6n_{\cal R}}$ ) the measure transforms as 
\begin{eqnarray}
[{\cal D}\chi][{\cal D}\lambda]\rightarrow e^{i\frac{2\pi k}{3}} [{\cal D}\chi][{\cal D}\lambda]\,.
\label{ 1 form 0 form mixed anomaly}
\end{eqnarray}
 The phase $e^{i\frac{2\pi}{3}}$  captures the $0$-form/$1$-form mixed anomaly. The general lore is that this mixed anomaly can be saturated in the IR by (1) massless excitations (as in the UV), (2) degenerate ground states, (3) topological quantum field theory (TQFT).  The saturation of the anomaly by $3$ degenerate ground states is manifest on $\mathbb R^3\times \mathbb S^1$, as we explicitly showed in the previous section. 
 
 As an exercise in group theory, we can also generalize the above construction to a $SU(N)$ gauge theory with fermions in several mixed representations, given that the theory is anomaly free and UV complete. We consider $n_1$ flavors of left-handed Weyl fermions in representation ${\cal R}_1$ and n-ality ${\cal N}_1$, $n_2$ fermions in representation ${\cal R}_2$ and n-ality ${\cal N}_2$, etc. The center symmetry of the theory is $\mathbb Z_p \subseteq \mathbb Z_N$. Since the n-ality of a representation  cannot change by adding gluon fields (they carry $0$ n-ality), we expect that the  center symmetry will depend only on the n-ality of a representation. Indeed, one can show that the maximum subgroup of the $1$-form center symmetry under which the theory with mixed representations  is invariant is $\mathbb Z_{p}^{(1)}$ with \cite{Cohen:2014swa}
 \begin{eqnarray}
 p=\mbox{gcd}\left(N,{\cal N}_1,{\cal N}_2,...\right)\,,
 \label{GCD in general}
 \end{eqnarray}
and gcd stands for the greatest common divisor. 
 This theory enjoys a set of global discrete chiral symmetries $\mathbb Z_{n_1 T({\cal R}_1)}\times\mathbb Z_{n_1 T({\cal R}_2)}\times...$, etc.   Now, we gauge $Z_{p}^{(1)}$ by turning on 't Hooft twists to find that under a discrete rotation the partition function receives an overall phase (assuming that the minimum fractional topological charger is $\frac{1}{p}$):
 \begin{eqnarray}
 \lambda_{{\cal R}_i}\rightarrow e^{i\alpha_i} \lambda_{{\cal R}_i}\implies \left[{\cal D} \lambda_{{\cal R}_i} \right]\rightarrow e^{i\frac{2\pi k}{p}}\left[{\cal D} \lambda_{{\cal R}_i} \right]\,,
 \label{general transformation}
 \end{eqnarray}
 where $k=0,2,..,p-1$ for each  representation, implying an anomaly between the $1$-form center symmetry and the corresponding   $0$-form discrete chiral symmetry.  The anomaly can be matched  by a CFT. Alternatively, it can be matched  by a $p$-degenerate ground state given that $\frac{n_i T({\cal R}_i)}{p}$ is an integer\footnote{Since there is a good $\mathbb Z_2$ symmetry for any Lorentz invariant theory, the discrete chiral symmetry cannot break to unity.}  $\geq 2$ for all representations. In this case the discrete symmetries break as
 \begin{eqnarray}
 \mathbb Z_{n_i T({\cal R}_i)}\rightarrow \mathbb Z_{\frac{n_i T({\cal R}_i)}{p}}\,,
 \end{eqnarray}
 for $i=1,2,...$.  One can trivially check that (\ref{GCD in general}) and (\ref{general transformation}) give the special  result (\ref{ 1 form 0 form mixed anomaly}) and the correct symmetry breaking pattern (\ref{the breaking pattern}).
  
Now if we take one of the $\mathbb T^4$ cycles to be smaller than $\Lambda$ a phase transition may occur. Nevertheless, the construction we adopted, 't Hooft twists, to find the $1$-form/$0$-form mixed anomaly still holds. This is why the anomaly can be used  both in the small- and large- circle limits to put constraints on the theory.    
 
In addition to the $1$-form 't Hooft anomaly, the theory admits various $0$-form (traditional) 't Hooft anomalies. In the next sections we study these anomalies for various numbers of adjoint and 3-index antisymmetric fermions,  examine their saturation in the IR, and draw conclusions about the  adiabatic continuity (if it exists) between theories defined on $\mathbb R^3\times \mathbb S_L^1$ and $\mathbb R^4$.

\subsection*{The IR spectrum for $n_G=0$ and $n_{\cal R}=1$}

In this case we have a single $0$-form 't Hooft  anomaly between the global $\mathbb Z^{\cal R}_6$ symmetry and a gravitational background ${\cal G}$, i.e., $\left[\mathbb Z_{6}^{\cal R}\right][{\cal G}]^2$ anomaly. Taking the charge of $\lambda$ to be $1$ (mod $ 6$) under $\mathbb Z_{6}^{\cal R}$, we find that  the coefficient of $\left[\mathbb Z_{6}^{\cal R}\right][{\cal G}]^2$ anomaly in the UV is $1\times d({\cal R})$, where $d({\cal R})=20$ is the dimension of the $3$-index antisymmetric representation. 
 
One option to match a $0$-form 't Hooft anomaly in the IR is to invoke composite massless fermions. However, a theory with a single fermion in the ${\cal R}$ representation does not admit color-singlet spin-$\frac{1}{2}$ operators. To prove this statement we first notice that building a spin-$\frac{1}{2}$ operator requires an odd number $2k+1$, $k \in \mathbb Z$, of fermions, each  carries  three indices $\lambda^{i_1i_2i_3}$. In total we have $3(2k+1)$ indices. Next, in order to form a color-singlet operator we need to contract the indices with $m$  Levi-Civita tensors $\epsilon_{i_1i_2...i_6}$, each carries $6$ indices. Contracting all the indices means that we have to satisfy the relation $3(2k+1)=6m$, which does not have a solution for any integers $k,m$. Next, we may want to build the composite fermions using $\lambda^{i_1i_2i_3}$ and the glue field $F_{i\,MN}^j$, which is valued in the adjoint representation and carries two color indices. We try to achieve this by using  $2k+1$  fermions and $\ell$ gluon fields. The indices can be contracted  using $m$ Levi-Civita tensors and $n$  Kronecker deltas $\delta_i^j$. Now, absorbing all the indices to form a color-singlet means that we need to satisfy the relation $3(2k+1)+2\ell=6m+2n$, which again does not have a solution for any integers $k,\ell,m,n$. We conclude that our theory does not admit  composite color-singlet fermionic operators.

The second option to match the anomaly is that the theory flows to a conformal point in the IR. However, this scenario is very implausible given the low dimensionality of ${\cal R}$, $d({\cal R})=20$. For example, a theory with a single adjoint fermion (super Yang-Mills), which has dimension $d(G)=35$, lies well outside the conformal window, see, e.g., the lattice study in \cite{Bergner:2015adz}. Therefore, it is very unlikely that a theory with a lower-dimensional fermion would flow to a conformal point. This conclusion is also supported by the $2$-loop $\beta$-function, which does not develop a fixed point in the case of a single fermion in the $3$-index antisymmetric representation. 

This leaves us with the last option: the discrete symmetry $\mathbb Z^{\cal R}_6$ has to break spontaneously in order for the  $\left[\mathbb Z_{6}^{\cal R}\right][{\cal G}]^2$ anomaly to be saturated. The natural question, then, is how does one detect the $\mathbb Z^{\cal R}_6$ breaking? As we noted above, the fermion bilinear 
$\epsilon_{i_1i_2...i_6}\epsilon_{\alpha\beta}\lambda^{\alpha\, i_1i_2i_3}$ $\times\lambda^{\beta\,i_4i_5i_6}$ vanishes identically. The next-to-leading operator (dimension $5$) is\footnote{Using covariant derivatives between a fermion bilinear does not help, since $D^M\lambda D_M\lambda=0$ for the exact same reason the fermion bilinear vanishes.} 
\begin{eqnarray}
{\cal O}_1=\epsilon_{i_4i_3j_1i_2i_5i_6}\epsilon_{\gamma\beta}\lambda^{\alpha\,i_1i_4i_3}F_{i_1\,MN}^{j_1}\left(\sigma^{MN}\right)_\alpha^\gamma\lambda^{\beta\,i_2i_5i_6}\,.
\label{possible condensate}
\end{eqnarray}
This operator is charged under $\mathbb Z_6^{\cal R}$ and can be used to detect the breaking $\mathbb Z_6^{\cal R}\rightarrow \mathbb Z_2$. The unbroken $\mathbb Z_2$ discrete symmetry is the $(-1)^F$ fermion number, which cannot be further broken in a Lorentz invariant theory. This leaves us with $3$ degenerate ground states, which is also compatible with the required number of ground states that match the mixed $1$-form/$0$-form 't Hooft anomaly. The four-fermion (dimension-$6$)  operator $\epsilon_{\alpha\beta}\epsilon_{\alpha'\beta'}\epsilon_{IJ}\epsilon_{I'J'}\lambda^{\alpha\,I}\lambda^{\alpha'\,J}\lambda^{\beta\,I'}\lambda^{\beta'\,J'}$, where $I,J$ are short-hand notations for $I\equiv i_1i_2i_3$, can also be used to detect the breaking $\mathbb Z_6^{\cal R}\rightarrow \mathbb Z_2$, since it carries $4$ charges under $\mathbb Z_6^{\cal R}$.

The absence of massless excitations  and the breaking pattern  $\mathbb Z_6^{\cal R}\rightarrow \mathbb Z_2$ in both $\mathbb R^3\times \mathbb S^1_L$ and $\mathbb R^4$  is a strong indication that the theory does not exhibit a phase transition as we decompactify $\mathbb S_L^1$, i.e., there is an adiabatic continuity from the small- to large- circle limits.

It is also instructive to track the mass gap as a function of the compactification radius. Using (\ref{photon mass for nR1 and nG0}) and expressing the mass gap in terms of the strong scale (with the aid of the one-loop $\beta$-function) we find:
\begin{eqnarray}
{\cal MG}_1(L)\sim\Lambda\left(\Lambda L\right)^{\frac{\beta_0}{4}-1}\,,\quad {\cal MG}_2(L)\sim\Lambda\left(\Lambda L\right)^{\frac{3\beta_0}{2}-1}\,,
\end{eqnarray}
where $\beta_0=20$ for $n_G=0,n_{\cal R}=1$. Therefore, the mass gap is a monotonically increasing function of $L$ such that at $L\gtrsim \Lambda^{-1}$ the theory enters its strongly coupled regime: the theory continues to confine the electric charges on $\mathbb R^4$. However, we expect a smooth transition from weak to strong coupling. 

\subsection*{The IR spectrum for $n_G=0$ and $n_{\cal R}>1$}

The global symmetry in this case is 
\begin{eqnarray}
\frac{SU(n_{\cal R})\times \mathbb Z^{\cal R}_{6n_{\cal R}}}{\mathbb Z_{n_{\cal R}}}\times \mathbb Z_{3}^{(1)}
\end{eqnarray}
and there are a few $0$-form anomalies that have to be matched between the UV and IR. We start by computing the anomalies in the UV.  As before, we take the charge of $\lambda$ to be $1$ under $\mathbb Z^{\cal R}_{6n_{\cal R}}$. The anomaly $\left[\mathbb Z_{6n_{\cal R}}^{\cal R}\right][{\cal G}]^2$ gives $n_{\cal R}\times d({\cal R})=20n_{\cal R}$. Also we can compute the anomaly $\left[\mathbb Z_{6n_{\cal R}}\right]\left[SU(n_{\cal R})\right]^2$ by examining the number of the fermion zero modes in the background of a $SU(n_{\cal R})$ BPST  instanton, which amounts to gauging $SU(n_{\cal R})$. This gives $1\times d({\cal R})=20$, where $1$ is the number of the fundamental Weyl zero modes in the background of the instanton (remember that $\lambda$ transforms as a fundamental in $SU(n_{\cal R})$). In addition, we have the anomaly $\left[SU(n_{\cal R}) \right]^3$, which is given by the cubic Dynkin index computed in the fundamental representation. In the UV this gives $d({\cal R})=20$. 

As in the $n_G=0,n_{\cal R}=1$ case, one cannot build gauge-invariant spin-$\frac{1}{2}$ composite operators: it is impossible to contract the indices of an odd number of $\lambda$, each carrying $3$ indices, to form a color singlet. Therefore, we expect that the anomalies in the IR will be matched either by a CFT, if the theory flows to a fixed point \cite{Banks:1981nn}, or that $SU(n_{\cal R})$ and  $\mathbb Z_{6n_{\cal R}}^{\cal R}$ will break spontaneously. The two-loop $\beta$-function (also see Table \ref{classification}) reveals that the theory develops a fixed point at weak coupling, $\alpha^*<1$, in the window $6\leq n_{\cal R}\leq 10$. In particular, we find that $\alpha^*\ll1$ at the upper end of the window, strongly suggesting that a theory with a large number of $\cal R$ flavors flows to a CFT.  

In order to study the breaking of  $SU(n_{\cal R})$ and  $\mathbb Z_{6n_{\cal R}}^{\cal R}$, which is expected to occur for $1<n_{\cal R}\leq 4$, see Table \ref{classification}, we need to build operators that transform nontrivially under the respective groups.   Let us start with the following operator
\begin{eqnarray}
{\cal O}_2^{qq'}=\epsilon_{IJ}\epsilon_{\alpha\beta}\lambda^{\alpha I q}\lambda^{\beta J q'}\,,
\label{Operator for chiSB}
\end{eqnarray}
which satisfies ${\cal O}_2^{qq'}=-{\cal O}_2^{q'q}$, where $q,q'$ are the flavor indices of $SU(n_{\cal R})$.  Hence, ${\cal O}_2^{qq'}$ has $n_{{\cal R}}(n_{\cal R}-1)/2$ components. A nonvanishing $\langle {\cal O}_2^{qq'}\rangle$ implies that $n_{\cal R}(n_{\cal R}-1)/2$ out of the $n_{\cal R}^2-1$ generators  of $SU(n_{\cal R})$ are broken (Goldstone bosons), leaving  $n_{\cal R}(n_{\cal R}+1)/2$ unbroken generators. Thus, the operator ${\cal O}_2^{qq'}$ can be used to examine the breaking of $SU(n_{\cal R})$ to a symplectic subgroup \cite{Li:1973mq}.

Next, we construct the color- and flavor-singlet operator:
\begin{eqnarray}
{\cal O}_3=\epsilon_{q_1q_2...q_{n_{\cal R}}}\epsilon_{q'_1q'_2...q'_{n_{\cal R}}} {\cal O}_2^{q_1q'_1} {\cal O}_2^{q_2q'_2}...{\cal O}_{2}^{q_{n_{\cal R}}q'_{n_{\cal R}}}\,.
\end{eqnarray}
This operator transforms under $\mathbb Z^{\cal R}_{6n_{\cal R}}$ as ${\cal O}_3\rightarrow e^{i\frac{2\pi}{6n_{\cal R}}(2n_{\cal R})}{\cal O}_3=e^{i\frac{2\pi}{3}}{\cal O}_3$, and hence, it can probe the $3$ degenerate ground states of the theory.  

Finally we comment on the mass gap on $\mathbb R^3\times \mathbb S^1_L$ and the behavior of the theory in the decompactification limit. Using (\ref{mass gap for nRG1 and nG0}) and the one-loop $\beta$-function, we find
\begin{eqnarray}
{\cal MG}_1\sim \Lambda\left(\Lambda L\right)^{\frac{9-n_{\cal R}}{2}}\,,\quad {\cal MG}_2\sim \Lambda\left(\Lambda L\right)^{10-n_{\cal R}}\,.
\label{mass gap for general r fermions}
\end{eqnarray}
Therefore, the mass gap monotonically increases with $L$, even for the maximum number of allowed  ${\cal R}$ fermions \footnote{Since ${\cal MG}_2>{\cal MG}_1$ for $L>\Lambda^{-1}$, we should always use ${\cal MG}_2$ to read the behavior of the mass gap in the decompactification limit.  Also, notice that  we need the next-to-leading order analysis to decide on the case $n_{\cal R}=10$.}. Contrasting  this result with the conformal window depicted in Table \ref{classification}, we see that for $n_{\cal R}\in [1,4]$ the theory flows to a strongly coupled regime as we increase the circle size. At values of $L\gtrsim \Lambda^{-1}$ we should stop trusting (\ref{mass gap for general r fermions}): the theory breaks its continuous chiral symmetry spontaneously (giving rise to Goldstone bosons)  to match its UV $0$-form 't Hooft anomalies. Yet, the theory cannot restore its broken discrete chiral symmetry, even in the strong coupling regime, since the $1$-form 't Hooft anomaly has to be matched. Thus, we expect that the theory will have $3$-degenerate ground states for all values of $L$. Also, the theory confines for all values of $L$. 

On the other hand, when $n_{\cal R}\in [5,10]$, then we expect from the two-loop $\beta$ function that the theory might flow to a CFT in the $L\rightarrow \infty$ limit. This conclusion is less robust at the lower corner of this window, thanks to the large coupling constant when $n_{\cal R}\cong 5$. Flowing to a CFT means that the theory has to restore its broken discrete chiral symmetry in the decompactification limit.  In all circumstance, one cannot trust  (\ref{mass gap for general r fermions}) for any number of fermions as long as $L\gtrsim \Lambda^{-1}$, since this expression was derived under the assumption that the W-boson mass $\frac{2\pi}{NL}$ is much larger than the photon mass $\sim e^{-S_m}/L$. This hierarchy of scales is lost when   $L\gtrsim \Lambda^{-1}$ and one can no longer trust the semi-classical analysis. If the theory flows to a CFT in the decompactification limit, then the mass gap has to turn from an increasing to a decreasing function of $L$ around $L\sim\Lambda^{-1}$.

We conclude that whether the theory flows to a CFT or breaks its continuous chiral symmetries, in both cases the spectrum on $\mathbb R^4$ is dramatically different from the spectrum of the theory on $\mathbb R^3\times \mathbb S^1_L$, which is fully gapped. Thus, we expect that the theory will experience a phase transition on the way from  $\mathbb R^3\times \mathbb S_L^1$ to $\mathbb R^4$.

\subsection*{The IR spectrum for $n_G=1$ and $n_{\cal R}=1$}

The global symmetries in this case are:
\begin{eqnarray}
U_{G-{\cal R}}(1)\times \mathbb Z_6^{\cal R}\times \mathbb Z_{12}^G\times \mathbb Z_3^{(1)}\,.
\end{eqnarray}
 The charges of the fermions under $U(1)_{G-{\cal R}}$ are taken to be $Q_{G}=1$ and $Q_{\cal R}=-2$, while the charges under the discrete symmetries are manifest. The $0$-form anomalies of this theory are listed in  Table \ref{Anomalies for nR1 and nG1}. 
  
\begin {table}
\begin{center}
\tabcolsep=0.11cm
\footnotesize
\begin{tabular}{|c|c|c|}
\hline
Anomaly & UV & IR\\
\hline
\hline
$\left[U_{G-{\cal R}}(1)\right]^3$ &  $d(G)Q_G^3+d({\cal R})Q_{{\cal R}}^3=-125$ & $\sum_{L=1,k_G=1,k_{\cal R}=0}L[k_G,k_{\cal R}]\left(2k_G+1-4k_{\cal R}\right)^3$ \\
\hline
$\left[U_{G-{\cal R}}(1)\right]\left[{\cal G}\right]^2$ & $d(G)Q_G+d({\cal R})Q_{{\cal R}}=-5$ & $\sum_{L=1,k_G=1,k_{\cal R}=0}L[k_G,k_{\cal R}]\left(2k_G+1-4k_{\cal R}\right)$\\
\hline
$\left[\mathbb Z_{6}^{{\cal R}}\right]\left[{\cal G}\right]^2$  &$d({\cal R})=20$ & $ \sum_{L=1,k_G=1,k_{\cal R}=0}L[k_G,k_{\cal R}]\left(2k_{\cal R}\right)$\\
\hline
$\left[\mathbb Z_{12}^{G}\right]\left[{\cal G}\right]^2$ & $d(G)=35$ & $ \sum_{L=1,k_G=1,k_{\cal R}=0}L[k_G,k_R]\left(2k_G+1\right)$\\
\hline
$\left[\mathbb Z_{12}^{G}\right]\left[U_{G-{\cal R}}(1)\right]^2$ & $d(G)Q_G^2=35$ & $\sum_{L=1,k_G=1,k_{\cal R}=0}L[k_G,k_{\cal R}](2k_G+1)\left(2k_G+1-4k_{\cal R}\right)^2$\\
\hline
$\left[\mathbb Z_{6}^{{\cal R}}\right]\left[U_{G-{\cal R}}(1)\right]^2$ & $d({\cal R})Q_{{\cal R}}^2=80$ & $\sum_{L=1,k_G=1,k_{\cal R}=0}L[k_G,k_{\cal R}](2k_{\cal R})\left(2k_G+1-4k_{\cal R}\right)^2$\\
\hline
$\left[\mathbb Z_{12}^{G}\right]^2\left[U_{G-{\cal R}}(1)\right]$  & $d(G)Q_G=35$ & $\sum_{L=1,k_G=1,k_{\cal R}=0}L[k_G,k_{\cal R}](2k_G+1)^2\left(2k_G+1-4k_{\cal R}\right)$   \\
\hline
$\left[\mathbb Z_{6}^{{\cal R}}\right]^2\left[U_{G-{\cal R}}(1)\right]$  & $d({\cal R})Q_{\cal R}=-40$ & $\sum_{L=1,k_G=1,k_{\cal R}=0}L[k_G,k_{\cal R}](2k_{\cal R})^2\left(2k_G+1-4k_{\cal R}\right)$   \\
\hline
$\left[\mathbb Z_{12}^{G}\right]^3$  & $d(G)Q_{G}^3=35$ & $\sum_{L=1,k_G=1,k_{\cal R}=0}L[k_G,k_{\cal R}]\left(2k_G+1\right)^3$   \\
\hline
$\left[\mathbb Z_{6}^{{\cal R}}\right]^3$  & $d({\cal R})Q_{{\cal R}}^3=20$ & $\sum_{L=1,k_G=1,k_{\cal R}=0}L[k_G,k_{\cal R}]\left(2k_{\cal R}\right)^3$   \\
\hline
$\left[\mathbb Z_{12}^{G}\right]\left[\mathbb Z_{6}^{{\cal R}}\right]^2$  & $0$ & $\sum_{L=1,k_G=1,k_{\cal R}=0}L[k_G,k_{\cal R}]\left(2k_{\cal R}\right)^2\left(2k_G+1\right)$   \\
\hline
$\left[\mathbb Z_{6}^{{\cal R}}\right]\left[\mathbb Z_{12}^{G}\right]^2$  & $0$ & $\sum_{L=1,k_G=1,k_{\cal R}=0}L[k_G,k_{\cal R}]\left(2k_{\cal R}\right)\left(2k_G+1\right)^2$   \\
\hline
\end{tabular}
\caption{\label{Anomalies for nR1 and nG1} The UV and IR anomalies for $n_G=1$ and $n_{\cal R}=1$. The IR and UV anomalies have to match, except for $\left[\mathbb Z_{N}\right]\left[U_{G-{\cal R}}(1)\right]^2$, where the anomalies have to match mod $N$ and for $\left[\mathbb Z_{N}\right]\left[{\cal G}\right]^2$, where the  UV and IR anomalies can differ by $mN+m'N/2$ and $m,m'$ are integers.  The last $6$ rows list  type II anomalies (according to the classification in \cite{Csaki:1997aw}) $\left[\mathbb Z_N\right]^2\left[U_{G-{\cal R}}(1)\right]$, $\left[\mathbb Z_N\right]^3$, and $\left[\mathbb Z_N\right]^2\left[\mathbb Z_M\right]$, where $N,M=6,12$. Here, the UV and IR anomalies can differ by $mN$ for $\left[\mathbb Z_N\right]^2\left[U_{G-{\cal R}}(1)\right]$, by  $mN+m'N^3/8$ for $\left[\mathbb Z_N\right]^3$, and by $m\,\mbox{gcd(N,M)}+m'N^2M/8$ for $\left[\mathbb Z_N\right]^2\left[\mathbb Z_M\right]$, where gcd is the greatest common divisor.}
\end{center}
\end{table}

We may attempt to match the $0$-form anomalies in the IR using composite fermions. A set of theses operators that are made of adjoint fermions are:
\begin{eqnarray}
\nonumber
\epsilon_{\alpha\gamma}\chi^{\alpha\,i}_j\chi^{\gamma\,j}_k\chi^{\beta \,k}_i\,,\quad F_{i\,MN}^j\left(\sigma^{MN}\right)^{\beta}_\alpha\chi^{\alpha\,i}_{j}\,,\quad F_{k\,MN}^j F_{iP}^{kN}\left(\sigma^{MP}\right)_\alpha^\beta\chi_{j}^{\alpha\, i}\,,\mbox{etc}.\\
\end{eqnarray}
In fact, we can dress any fermion $\chi$ with an arbitrary number of gluon operators, and in principle, this set of operators is unbounded\footnote{In fact, a bound on the number of such composites, which can be very large, come from the a-theorem \cite{Cardy:1988cwa}.}. 
Another set of fermion operators can be obtained by appending $2$ fermions in the $3$-index antisymmetric representation to any of the above operators. Actually, we can append only an even number of $\lambda$, otherwise there is no way of contracting the color indices, as we showed above. For example, we can write the operators
\begin{eqnarray}
\nonumber
&&\epsilon_{\alpha\rho}\epsilon_{\gamma\sigma}\epsilon_{i_1i_2i_3i_4i_5i_6}\chi^{\alpha\,i}_j\chi^{\gamma\,j}_k\chi^{\beta\,k}_i\lambda^{\rho\,i_1i_2i_3}\lambda^{\sigma\,i_4i_5i_6}\,,\\
&&\epsilon_{\gamma\rho}\epsilon_{i_1i_2i_3i_4i_5i_6}F_{j\,MN}^{k}\left(\sigma^{MN}\right)^{\beta}_\alpha\chi^{\gamma\,j}_k\lambda^{\alpha\,i_1i_2i_3}\lambda^{\rho\,i_4i_5i_6}\,,\mbox{etc}.\,.
\end{eqnarray}

We generalize this construction as follows. Let $L(k_G,k_{\cal R})$ be the number of fermion operators that have $2k_G+1$ insertions of $\chi$ and $2k_{\cal R}$ insertions of $\lambda$, where $k_G,k_{\cal R}\in \mathbb Z^+$. $L(k_G,k_{\cal R})$ can be chosen to be any positive integer by dressing a given number of $G$ and $\cal R$ fermions with gluon fields.  The IR anomalies of these operators are listed in the third column of Table \ref{Anomalies for nR1 and nG1}.

In principle, there is an infinite number of IR scenarios that can match the anomalies. An economical choice is given by the union of the following two sets of fermions\footnote{We have also checked that these composites match the $0$-form anomalies of the discrete subgroups $\mathbb Z^{G}_4\subset \mathbb Z_{12}^G$ and $\mathbb Z^{{\cal R}}_2\subset \mathbb Z^{{\cal R}}_{6}$.}
\begin{eqnarray}
\nonumber
\mbox{set 1 with}&& L=1, k_G=2\,, k_{\cal R}=0: \quad\chi\chi\chi\chi\chi\,,\\
\mbox{set 2 with}&&  L=2, k_G=1\,, k_{\cal R}=2:\quad \chi\chi\chi\lambda\lambda\lambda\lambda\,, F\chi\chi\chi\lambda\lambda\lambda\lambda\,. 
\label{set of IR operators for ng and nr 1}
\end{eqnarray}
However, we expect that the $1$-form/$0$-form anomaly to be matched by a TQFT. It is an interesting question to investigate the nature of this TQFT along the lines of \cite{Wan:2018bns}, where TQFT of QCD with adjoint fermions were constructed. 

We could also repeat the same exercise above by building operators of the form
\begin{eqnarray}
\epsilon^{\dot\alpha\dot\beta}\bar \lambda_{ijk\,\dot\alpha}\bar\chi_{l \dot\beta}^j\lambda^{ikl\,\beta}\,,...,\mbox{etc}.\,.
\end{eqnarray}

The  anomalies can also be saturated  in the IR by the breaking of  $U_{G-{\cal R}}(1)$, $\mathbb Z_{12}^{G}$, and $\mathbb Z_{6}^{\cal R}$.   In the following,  we examine the operators that probe this breaking. The lowest order operator, which is charged under $U_{G-{\cal R}}(1)$,  reads:
\begin{eqnarray}
{\cal O}_4=\epsilon_{\alpha\beta}\chi^{\alpha\,i}_{ j}\chi^{\beta\,j}_{i}\,.
\end{eqnarray}
Another operator that transforms under all the symmetries is
\begin{eqnarray}
{\cal O}_5=\epsilon_{IJ}\epsilon_{\alpha\beta}\epsilon_{\gamma\delta}\chi^{\alpha j}_{i}\chi^{\gamma i}_{ j}\lambda^{\beta I}\lambda^{\delta J}\,.
\end{eqnarray}
A non-zero vacuum expectation value of ${\cal O}_4$ or ${\cal O}_5$ indicates the breaking of $U_{G-{\cal R}}(1)$. 
One can also construct a $U_{G-{\cal R}}(1)$ invariant operator, which is charged under the various discrete groups:
\begin{eqnarray}
{\cal O}_6=\epsilon_{\alpha_1\alpha_5}\epsilon_{\alpha_2\alpha_6}\epsilon_{\alpha_3\alpha_4}\epsilon_{IJ}\chi^{\alpha_1\,i_1}_{i_2}\chi^{\alpha_2\,i_2}_{i_3}\chi^{\alpha_3\,i_3}_{i_4}\chi^{\alpha_4\,i_4}_{i_1}\lambda^{\alpha_5 I}\lambda^{\alpha_6 J}\,.
\end{eqnarray}
This operator transforms as ${\cal O}_6\rightarrow e^{i\frac{2\pi}{3}}{\cal O}_6$ under the  discrete groups. A nonvanishing vacuum expectation values  of this operator signals the breaking  $\mathbb Z_6^{\cal R}\rightarrow \mathbb Z_2$ and $\mathbb Z_{12}^G\rightarrow \mathbb Z_4$, which is also compatible with the $1$-form/$0$-form mixed anomaly.

Now, we discuss the connection to the spectrum on $\mathbb R^3\times \mathbb S^1_L$. We remind the reader that the vacuum is $3$-fold degenerate and there is a single massless excitation in the IR. This is one of the dual photons, which can be thought of as the Goldstone boson resulting from the breaking of $U_{G-{\cal R}}(1)$.  The mass gaps of this theory are given by
\begin{eqnarray}
{\cal MG}_1\sim \Lambda \left(L\Lambda\right)^{7}\,, \quad {\cal MG}_2\sim \Lambda \left(L\Lambda\right)^{-25}\,.
\end{eqnarray}
Since ${\cal MG}_1$ is a monotonically increasing function of $L$, the semi-classical analysis will break when $L\gtrsim \Lambda^{-1}$. Then, how does the theory behave in the decompactification limit? The two-loop $\beta$-function does not have a fixed point for $n_G=1$ and $n_{\cal R}=1$, and therefore, it is less likely that the theory flows to a CFT in the IR. One plausible scenario, which is adiabatically connected to the theory on $\mathbb R^3\times \mathbb S^1_L$, is that  $U_{G-{\cal R}}(1)$ remains broken. In fact, the operator ${\cal O}_4$ that gets a vacuum expectation value and signals the breaking of $U_{G-{\cal R}}(1)$ on $\mathbb R^4$ is the exact same operator that signal the same breaking on $\mathbb R^3\times \mathbb S^1_L$.  In addition, the vacuum on $\mathbb R^4$ remains $3$-fold degenerate.  This hints that theories on $\mathbb R^4$ and  $\mathbb R^3\times \mathbb S^1_L$ might be continuously connected. 

Yet, another scenario is that $U_{G-{\cal R}}(1)$ is restored in the decompactification limit and the anomalies in the IR are matched by the operators (\ref{set of IR operators for ng and nr 1}) along with a TQFT.

\subsection*{The IR spectrum for the general case $n_G >1$ and $n_{\cal R}\geq1$}

Having more than one adjoint fermion makes the story more complicated since the number of the anomalies one needs to match increases dramatically. Here, we consider the simplest case of $n_G=2$ and $n_{\cal R}=1$. In particular, we would like to answer the question of adiabatic continuity between $\mathbb R^3\times \mathbb S^1$ and $\mathbb R^4$. We recall that the theory on $\mathbb R^3\times \mathbb S^1$ has one scalar and one fermionic (in the fundamental of $SU(n_G=2)$) degrees of freedom.  The  UV theory has $\left[SU(n_G)\right]^3$ anomaly (of Witten's type), and therefore, it has to be matched in the IR on $\mathbb R^4$ by an odd number of fermion operators. The theory also has a $\left[U_{G-{\cal R}}(1)\right]^3$ anomaly, which gives in the UV $n_GQ_G^3d(G)+Q_{\cal R}^3d({\cal R})=-90$,   $\left[U_{G-{\cal R}}\right][{\cal G}]^2$, which gives $n_{G}d(G)Q_G+d({\cal R})Q_{\cal R}=30$, and $\left[U_{G-{\cal R}}(1)\right]\left[SU(n_G)\right]^2$, which gives  $d(G)Q_G=35$. In addition, there are various anomalies between the  discrete symmetries $\mathbb Z_{12n_G}^G$, $\mathbb Z^{{\cal R}}_{6n_{\cal R}}$ and  $U_{G-{\cal R}}(1)$, $SU(n_G)$, and ${\cal G}$.

Let us assume that these anomalies can be matched in the IR by a set of composite fermions. Since these operators are made of an odd number of adjoint fermions, they have to transform in odd representations of $SU(n_G=2)$: $(1), (3),(5),...,$ etc. A set of {\em skeleton operators} (made only of adjoints) is given by
\begin{eqnarray}
\underbrace{\chi\chi\chi}_{(1)}\, \quad \underbrace{\chi\chi\chi}_{(3)}\,,\quad \underbrace{\chi\chi\chi\chi\chi}_{(1)}\,,\quad \underbrace{\chi\chi\chi\chi\chi}_{(3)}\,,\quad \underbrace{\chi\chi\chi\chi\chi}_{(5)}\,,...\mbox{etc.}\,,
\end{eqnarray}
where the underbrace is the $SU(2)$ representation they transform in. Each of these operators can be dressed with mutiple gluon fields and/or fermions in the $3$-index antisymmetric representation. We take $L[k_G,k_{\cal R};R_G]$ to denote the number of operators that have $2k_G+1$ insertions of adjoints, $2k_{\cal R}$ insertions of ${\cal R}$ fermions, such that these operators transform in the representation $R_G$ of $SU(n_G=2)$ (only odd representations appear). Then, a subset of 't Hooft anomalies in the IR are given by
\begin{eqnarray}
\nonumber
\left[U_{G-{\cal R}}(1)\right]^3&=&\sum_{L=1,k_G=1,k_{\cal R}=0, R_G=1} \mbox{dim}(R_G)L[k_G,k_{\cal R};R_G](2k_G+1-4k_{\cal R})^3\,,\\
\nonumber
\left[U_{G-{\cal R}}(1)\right]\left[{\cal G}\right]^2&=&\sum_{L=1,k_G=1,k_{\cal R}=0,R_G=1} \mbox{dim}(R_G)L[k_G,k_{\cal R};R_G](2k_G+1-4k_{\cal R})\,,\\
\left[U_{G-{\cal R}}(1)\right]\left[SU(n_G)\right]^2&=&\sum_{L=1,k_G=1,k_{\cal R}=0,R_G=1}T(R_G)L[k_G,k_{\cal R};R_G](2k_G+1-4k_{\cal R})\,,
\label{IR anomalies for nr and ng 1}
\end{eqnarray} 
where $T(R_G)$ is the trace operator defined by $\mbox{tr}_{R_G}\left[T^aT^b\right]=\delta^{ab}T(R_G)$, and $\{T^a\}$ are the Lie generators of $SU(2)$. We were not able to find a set of composite fermions with a relatively small $L, k_G,k_{\cal R}$ that could match all the anomalies. Also, we could not  prove that (\ref{IR anomalies for nr and ng 1}) has no solution. We conclude that if a set of composite fermions can be found, it must have a nontrivial combination of higher-dimensional operators. This is an exercise in number theory that we leave for future studies. Otherwise, the anomaly can be matched either by a CFT (we see from Table \ref{classification} that this is a possibility, not very robust though) or that  $U_{G-{\cal R}}(1)$ and $SU(n_G)$ will break spontaneously. 

As we increase the number of adjoint or $3$-index antisymmetric fermions the anomaly matching conditions become more and more clumsy. We  expect that as the number of fermions increases the theory flows to a CFT, thus matching the anomalies is trivialized. However, in all cases the semi-classical analysis on the circle will break at scales $L\gtrsim \Lambda^{-1}$ since the mass gap on the circle $\sim\Lambda (L\Lambda)^{10-2n_G-n_{\cal R}}$ is a monotonically increasing function of $L$ for all $n_G$ and $n_{\cal R}$. We also expect a phase transition will occur as we transit from $\mathbb R^3\times \mathbb S^1_L$ to $\mathbb R^4$.

\section{Outlook}
\label{outlock}

\begin {table}
\begin{center}
\tabcolsep=0.11cm
\footnotesize
\begin{tabular}{|c|c|c|}
\hline
$(n_G,n_{\cal R})$ & IR symmetry realization & Adiabatic continuity to $\mathbb R^3\times \mathbb S^1_L$? \\
\hline\hline
$(0,1)$ & $ \slashed{\mathbb Z}_6^{\cal R}$ & $\checkmark$\\
\hline
$(0,[1,4])$ & $\slashed{\mathbb Z}_{6n_{\cal R}}^{\cal R}$, $\slashed{SU}(n_{\cal R})$  & $\times$\\
\hline
$(0,[5,10])$ & CFT  & $\times$\\
\hline
$(1,1)$ & $\slashed{U}_{G-{\cal R}}(1)$, $\slashed{\mathbb Z}_{6}^{\cal R}$, $\slashed{\mathbb Z}_{12}^{G}$, OR & $\checkmark$\\
 & composite fermions+TQFT & $\times$\\
\hline
$(2,1)$ &  $\slashed{\mathbb Z}_{6}^{\cal R}$, $\slashed{U}_{G-{\cal R}}(1)$, $\slashed{\mathbb Z}_{24}^{G}$, $\slashed{SU}(n_{G}=2)$ & $\times$\\
\hline
\end{tabular}
\caption{\label{outlook on all cases}A summary of the IR symmetry realization of the theory on $\mathbb R^4$ for the main cases we considered in this paper.}
\end{center}
\end{table}

In this paper we carried out an in-depth  analysis of  QCD with fermions in the adjoint and the $3$-index antisymmetric representations.  Our main task was to study the possible dynamical realizations of the $UV$ symmetries. To achieve this task, we used semi-classical analysis and 't Hooft anomaly matching conditions and draw a few possible non-trivial connections between the spectrum on the circle and on $\mathbb R^4$. The main results are summarized in Table \ref{outlook on all cases}.

Our major conclusions and lessons can be grouped as follows:
\begin{enumerate}

\item Yang-Mills theory with a single fermion in the $3$-index antisymmetric representation does not exhibit a phase transition as we dial the circle size $L$. There is an adiabatic continuity as $L$ interpolates from small to large sizes. The ground state is $3$-fold degenerate, the theory is gapped, and it confines the electric charges for all values of $L$.

\item Yang-Mills theory with two or more fermions in the $3$-index antisymmetric representation will exhibit a phase transition as $L$ is dialed across $\Lambda^{-1}$. In the small circle limit the theory is fully gapped, the continuous chiral symmetry is intact, the theory is in the confined phase, and the ground state is $3$-fold degenerate, meaning that the discrete chiral symmetry is broken. In the large circle limit the theory is still confining, but otherwise it breaks its continuous chiral symmetry leading to massless Goldstone bosons in the spectrum. Yet, the ground state is still $3$-fold degenerate and the discrete symmetries are broken. In this regard, one wonders what kind of boundary conditions one might apply on the small circle in order to avoid a phase transition in the decompactification limit. One plausible way to meet this requirement is to force the theory to break its continuous chiral symmetry in the small circle limit along the lines of \cite{Cherman:2016hcd}. Such a possibility is left for future studies.  

\item There is a possible adiabatic continuity  when we have a single adjoint fermion and another fermion in the $3$-index antisymmetric representation. In this case there is a global $U_{G-{\cal R}}(1)$ symmetry, which remains broken as we dial the circle size. Again, the ground state is $3$-fold degenerate and the theory confines in the IR.  Interestingly, there is also another realization of the IR symmetries on $\mathbb R^4$, which is given by the set of composite fermions (\ref{set of IR operators for ng and nr 1}). In this case, the theory retain its $U_{G-{\cal R}}(1)$ symmetry intact and matches the $1$-form/$0$-form anomaly via a TQFT. Now, if we compactify this theory on a circle, give the composite fermions periodic boundary conditions, and take $L$ smaller that $\Lambda^{-1}$ we should expect that the global $U_{G-{\cal R}}(1)$ will break and a Goldstone boson, the massless dual photon, will appear. The puzzling thing, though, is that the massless fermions should also get gapped. This appearance/disappearance of massless fermionic degrees of freedom in a phase transition is not expected to happen. Resolving this puzzle calls for further investigation of this case. 

\end{enumerate}

{\flushleft \bf Acknowledgments:} The author thanks Erich Poppitz and Mithat \"Unsal for valuable discussions. This work is supported by NSF grant PHY-1720135. 
\bibliography{References}

\bibliographystyle{JHEP}

\end{document}